\PassOptionsToPackage{usenames,dvipsnames}{xcolor}
\PassOptionsToPackage{hyphens}{url}
\PassOptionsToPackage{showframe}{geometry}

\documentclass[conference]{IEEEtran}

\usepackage{caption}
\usepackage{ragged2e}
\usepackage[bookmarks=true,breaklinks=true,letterpaper=true,colorlinks,citecolor=blue,linkcolor=blue,urlcolor=blue]{hyperref}
\usepackage{cite}
\usepackage[numbers,sort&compress]{natbib}

\DeclareCaptionFont{specialsize}{\fontsize{8pt}{10pt}\selectfont}
\usepackage[captionskip=0pt]{floatrow}

\makeatletter
\newcommand{\linebreakand}{%
\end{@IEEEauthorhalign}
\hfill\mbox{}\par
\mbox{}\hfill\begin{@IEEEauthorhalign}
}
\makeatother

\author{
  \IEEEauthorblockN{Suyash Mahar}
  \IEEEauthorblockA{\textit{%
      UC San Diego}
    \\
  }
  \and
  \IEEEauthorblockN{Mingyao Shen}
  \IEEEauthorblockA{\textit{
      UC San Diego} \\
  }  \and
  \IEEEauthorblockN{Terence Kelly}
  \IEEEauthorblockA{\textit{Independent} %
  } \and
  \IEEEauthorblockN{Steven Swanson}
  \IEEEauthorblockA{\textit{
      UC San Diego }%
  }
}

\makeatletter                   
\def\mdseries@tt{m}             
\makeatother                    

\usepackage[plain]{fancyref}
\usepackage[draft=true]{minted} 
\usepackage{color}
\usepackage{textgreek}
\usepackage{breakurl}           

\usepackage{amssymb}
\usepackage{lipsum}
\usepackage[final]{pdfpages}
\usepackage{setspace}
\usepackage{epsf}
\usepackage{epsfig}
\usepackage{graphicx}
\usepackage{algorithmic}
\usepackage{amsmath}
\usepackage{letltxmacro}

\usepackage{enumitem}
\usepackage{tikz}
\usepackage{tightenum}
\usepackage{xspace}

\usepackage{url}
\usepackage{multirow}

  \renewenvironment{thebibliography}[1]{%
    \begin{oldthebibliography}{#1}%
      \setlength{\parskip}{0ex}%
      \setlength{\itemsep}{0ex}%
  }%
  {%
    \end{oldthebibliography}%
  }

\newwrite\arxivdeps
\immediate\openout\arxivdeps=\jobname-arxivdeps.log

\newcommand\verifymarkedforarxivfile[1]{%
\ifdefined\arxivbuild
\else
\IfFileExists{#1}%
{}%
{\GenericWarning{Marked file (#1) for inclusion in arxiv build does not exist}}
\fi%
}

\newcommand\markforarxiv[1]{%
\verifymarkedforarxivfile{#1}%
\write\arxivdeps{IncludeInArxiv: #1}%
}

\DeclareUrlCommand\UScore{\urlstyle{rm}}

\LetLtxMacro\oldincludegraphics\includegraphics
\renewcommand{\includegraphics}[2][]{%
\markforarxiv{#2}%
\oldincludegraphics[#1]{#2}}

\LetLtxMacro\oldincludepdf\includepdf
\renewcommand{\includepdf}[2][]{%
\markforarxiv{#2}%
\oldincludepdf[#1]{#2}}

\def\nfigure[#1,#2,#3]{
\begin{figure}
\vspace*{0mm}
\begin{center}

\includegraphics[width=\columnwidth]{#1} 
\caption[]{#2
} \label{#3}

\end{center}
\end{figure}}

\def\cfigure[#1,#2,#3]{
\begin{figure}
\vspace*{0mm}
\begin{center}

\includegraphics[width=3in]{#1} 
\caption[]{#2
} \label{#3}
 
\end{center}
\end{figure}}

\def\cfigurefour[#1,#2,#3]{
\begin{figure}
\vspace*{0mm}
\begin{center}

\includegraphics[width=4in]{#1} 
\vspace*{-3mm}\caption[]{#2
} \label{#3}
 
\vspace*{-5mm}
\end{center}
\end{figure}}

\def\cfiguretemp[#1,#2,#3]{
\begin{figure}
\vspace*{0mm}
\begin{center}

\includegraphics[width=3.5in]{#1} 
\caption[]{#2
} \label{#3}
 
\end{center}
\end{figure}}

\def\wfigure[#1,#2,#3]{
\begin{figure*}
\vspace*{0mm}
\begin{center}
 \includegraphics[width=\textwidth]{#1} 
 \caption[]{#2
} \label{#3}
 
\end{center}
\end{figure*}}

\def\threefigure[#1,#2,#3,#4,#5]{
\begin{figure*}
\vspace*{0mm}
\begin{center}

\begin{tabular}{ccc}
\includegraphics[width=2in]{#1} & \includegraphics[width=2in]{#2} &  \includegraphics[width=2in]{#3} \\
(a) & (b) & (c) \\
\end{tabular}
\vspace*{-3mm}\caption[]{#4
} \label{#5}

\vspace*{-5mm}
\end{center}
\vspace*{-2mm}
\end{figure*}}

\def\dcfigure[#1,#2,#3,#4,#5,#6]{
{
\begin{figure*}
\begin{center}
\begin{minipage}[c]{\columnwidth}{
\includegraphics[width=\columnwidth]{#1} 
\vspace*{0mm}\caption[]{#2} \label{#3} \
}\end{minipage}\hspace*{\columnsep}\
\begin{minipage}[c]{\columnwidth}{
\includegraphics[width=\columnwidth]{#4} 
\vspace*{0mm}\caption[]{#5}\label{#6} \
}\end{minipage}
\end{center}
\end{figure*}
}
}

\def\scfigure[#1,#2,#3]{
{
\begin{figure*}
\begin{center}
\begin{minipage}[c]{3.5in}{
\includegraphics[width=3.5in]{#1} 
}\end{minipage}
\caption[]{#2} \label{#3} \
\end{center}
\end{figure*}
}
}

\def\tableByTable[#1,#2,#3,#4,#5,#6]{
{
\begin{table*}
\begin{center}
\begin{minipage}[c]{3in}{
\centering
{#1}
\vspace*{0mm}\tabcaption[]{#2}\label{#3} \
}\end{minipage}\hspace*{\columnsep}\
\begin{minipage}[c]{3in}{
\centering
{#4}
\vspace*{0mm}\tabcaption[]{#5}\label{#6} \
}\end{minipage}
\end{center}
\end{table*}
}
}

\def\figureByTable[#1,#2,#3,#4,#5,#6]{
{
\begin{figure*}
\begin{center}
\begin{minipage}[c]{3in}{
\centering
\includegraphics[width=\textwidth]{#1}
\vspace*{0mm}\figcaption[]{#2} \label{#3} \
}\end{minipage}\hspace*{\columnsep}\
\begin{minipage}[c]{3.3in}{
\centering
{#4}
\vspace*{0mm}\tabcaption[]{#5}\label{#6} \
}\end{minipage}
\end{center}
\end{figure*}
}
}

\def\tableByFigure[#1,#2,#3,#4,#5,#6]{
{
\begin{figure*}
\begin{center}
\begin{minipage}[c]{4.3in}{
\centering
{#1}
\vspace*{0mm}\tabcaption[]{#2} \label{#3} \
}\end{minipage}\hspace*{\columnsep}\
\begin{minipage}[c]{2.2in}{
\centering
\includegraphics[width=\textwidth]{#4}
\vspace*{-0.35in}\caption[]{#5}\label{#6} \
}\end{minipage}
\end{center}
\end{figure*}
}
}

\def\doublecfigure[#1,#2,#3,#4]{
{
\begin{figure}
\begin{center}
\begin{minipage}[c]{1.5in}{
\begin{center}
\includegraphics[width=1.5in]{#1}
\end{center}
}\end{minipage}\hspace*{1em}\
\begin{minipage}[c]{1.5in}{
\begin{center}
\includegraphics[width=1.5in]{#2}
\end{center}
}\end{minipage}
\vspace*{0mm}\caption[]{#3} \label{#4} \
\end{center}
\end{figure}
}
}

\def\qcfigure[#1,#2,#3,#4,#5,#6]{
{
\begin{figure*}
\vspace*{0.2in}\
\begin{center}
\begin{minipage}[c]{3in}{
\includegraphics[width=3in]{#1} 
\vspace*{-3mm}
}
\end{minipage}\hspace*{0.5in}\
\begin{minipage}[c]{3in}{
\includegraphics[width=3in]{#2} 
\vspace*{-3mm}
}\end{minipage}

\begin{minipage}[c]{3in}{
\includegraphics[width=3in]{#3} 
\vspace*{-3mm}
}
\end{minipage}\hspace*{0.5in}\
\begin{minipage}[c]{3in}{
\includegraphics[width=3in]{#4} 
\vspace*{-3mm}
}\end{minipage}
\end{center}
\caption[]{#5}\label{#6}
\end{figure*}
}
}

\def\twfigure[#1,#2,#3,#4,#5]{
{
\begin{figure*}
\vspace*{0.2in}\
\begin{center}
\begin{minipage}[c]{6.5in}{
\includegraphics[width=6.5in]{#1} 
\vspace*{-3mm}
}
\end{minipage}

\begin{minipage}[c]{6.5in}{
\includegraphics[width=6.5in]{#2} 
\vspace*{-3mm}
}\end{minipage}

\begin{minipage}[c]{6.5in}{
\includegraphics[width=6.5in]{#3} 
\vspace*{-3mm}
}
\end{minipage}
\end{center}
\caption[]{#4}\label{#5}
\end{figure*}
}
}

\def\dwfigure[#1,#2,#3,#4]{
{
\begin{figure*}
\vspace*{0.2in}\
\begin{center}
\begin{minipage}[c]{6.5in}{
\includegraphics[width=6.5in]{#1} 
\vspace*{-3mm}
}
\end{minipage}

\begin{minipage}[c]{6.5in}{
\includegraphics[width=6.5in]{#2} 
\vspace*{-3mm}
}\end{minipage}

\end{center}
\caption[]{#3}\label{#4}
\end{figure*}
}
}

\def\dssfigure[#1,#2,#3,#4,#5,#6]{
{
\begin{figure*}
\vspace*{0.2in}\
\begin{center}
\begin{minipage}[c]{4in}{
\includegraphics[width=4in]{#1}
\vspace*{-3mm}\caption[]{#2} \label{#3} \
}\end{minipage}\hspace*{0.5in}\
\begin{minipage}[c]{2in}{
\includegraphics[width=2in]{#4}
\vspace*{-3mm}\caption[]{#5}\label{#6} \
}\end{minipage}
\end{center}
\vspace*{-0.4in}\
\end{figure*}
}
}

\def\dsfigure[#1,#2,#3,#4,#5,#6]{
{
\begin{figure*}
\vspace*{0.2in}\
\begin{center}
\begin{minipage}[c]{3in}{
\includegraphics[width=3in]{#1}
\vspace*{-3mm}\caption[]{#2} \label{#3} \
}\end{minipage}\hspace*{0.5in}\
\begin{minipage}[c]{3in}{
\hspace*{0.5in}\
\includegraphics[height=3in]{#4}
\vspace*{-3mm}\caption[]{#5}\label{#6} \
}\end{minipage}
\end{center}
\vspace*{-0.4in}\
\end{figure*}
}
}

\def\dsyfigure[#1,#2,#3,#4,#5,#6]{
{
\begin{figure*}
\vspace*{0.2in}\
\begin{center}
\begin{minipage}[c]{2.5in}{
\includegraphics[height=2.5in]{#1}
\vspace*{-3mm}\caption[]{#2} \label{#3} \
}\end{minipage}\hspace*{0.5in}\
\begin{minipage}[c]{2.5in}{
\includegraphics[height=2.5in]{#4}
\vspace*{-3mm}\caption[]{#5}\label{#6} \
}\end{minipage}
\end{center}
\vspace*{-0.4in}\
\end{figure*}
}
}

\def\dyfigure[#1,#2,#3,#4,#5,#6]{
{
\begin{figure*}
\vspace*{0.2in}\
\begin{center}
\begin{minipage}[c]{3in}{
\includegraphics[height=3in]{#1} 
\vspace*{-3mm}\caption[]{#2} \label{#3} \
}\end{minipage}\hspace*{0.5in}\
\begin{minipage}[c]{3in}{
\includegraphics[height=3in]{#4} 
\vspace*{-3mm}\caption[]{#5}\label{#6} \
}\end{minipage}
\end{center}
\vspace*{-0.4in}\
\end{figure*}
}
}

\def\dyoldfigure[#1,#2,#3,#4,#5,#6]{
{
\begin{figure*}
\vspace*{0.2in}\
\begin{center}
\begin{minipage}[c]{3in}{
\epsfysize=2.0in\
\hspace{0.5in}\
\epsfbox{#1}
\vspace*{-3mm}\caption[]{#2} \label{#3} \
}\end{minipage}\hspace*{0.25in}\
\begin{minipage}[c]{3in}{
\epsfysize=2.0in\
\hspace{0.5in}\
\epsfbox{#4}
\vspace*{-3mm}\caption[]{#5}\label{#6} \
}\end{minipage}
\end{center}
\vspace*{-0.4in}\
\end{figure*}
}
}

\def\cfiguredouble[#1,#2,#3,#4]{
\begin{figure}
\vspace*{0.2in}\
\begin{center}
\begin{minipage}[c]{1.5in}{
\epsfxsize=1.5in\
\epsfbox{#1}
}\end{minipage}\hspace*{0.1in}\
\begin{minipage}[c]{1.5in}{
\epsfxsize=1.5in\
\vspace{0.1in}\epsfbox{#2}
}\end{minipage}\vspace*{-0.10in} \caption[]{#3}\label{#4}
\end{center}
\vspace*{-0.4in}\
\end{figure}
}

\def\wpfigure[#1,#2,#3,#4]{
\begin{figure*}
\vspace*{4mm}
\begin{center}

\includegraphics[width=#4]{#1} 

\vspace*{-3mm}\caption[]{#2
} \label{#3}

\vspace*{-5mm}
\end{center}
\end{figure*}}

\def\wprfigure[#1,#2,#3,#4,#5]{
\begin{figure*}
\vspace*{4mm}
\begin{center}

\includegraphics[width=#4, angle=#5]{#1} 

\vspace*{-3mm}\caption[]{#2
} \label{#3}

\vspace*{-5mm}
\end{center}
\end{figure*}}

\def\DoubleFigureWSlide[#1,#2,#3,#4,#5,#6,#7,#8,#9]{
\begin{figure*}
\vspace*{#9}
\begin{center}
\begin{minipage}{#4}
\includegraphics[width=#4]{#1}
\vspace*{-3mm}\caption{#2
}\label{#3}
\end{minipage}
\hspace{2em}
\begin{minipage}{#8}
\includegraphics[width=#8]{#5}
\vspace*{-3mm}\caption{#6
}\label{#7}
\end{minipage}
\vspace*{-5mm}
\end{center}
\end{figure*}
}

\def\DoubleFigureW[#1,#2,#3,#4,#5,#6,#7,#8]{
\begin{figure*}
\vspace*{0in}
\begin{center}
\begin{minipage}{#4}
\includegraphics[width=#4]{#1}
\vspace*{-3mm}\caption{#2
}\label{#3}
\end{minipage}
\hspace{2em}
\begin{minipage}{#8}
\includegraphics[width=#8]{#5}
\vspace*{-3mm}\caption{#6
}\label{#7}
\end{minipage}
\vspace*{-5mm}
\end{center}
\end{figure*}
}

\def\DoubleFigureWHack[#1,#2,#3,#4,#5,#6,#7,#8]{
\begin{figure*}
\vspace*{0in}
\begin{center}
\begin{minipage}{3in}
\includegraphics[width=#4]{#1}
\vspace*{-3mm}\caption{#2
}\label{#3}
\end{minipage}
\hspace{2em}
\begin{minipage}{3in}
\includegraphics[width=#8]{#5}
\vspace*{-3mm}\caption{#6
}\label{#7}
\end{minipage}
\vspace*{-5mm}
\end{center}
\end{figure*}
}

\def\ddcfigure[#1,#2,#3,#4]{
\begin{figure*}
\vspace*{0.2in}\
\begin{center}
\begin{minipage}[c]{\columnwidth}{
\includegraphics[width=\columnwidth]{#1} 
}\end{minipage}\hspace{0.5in}\
\begin{minipage}[c]{\columnwidth}{
\includegraphics[width=\columnwidth]{#2} 
}\end{minipage} \caption[]{#3}\label{#4}
\end{center}
\end{figure*}
}

\def\ddcfigureSlide[#1,#2,#3,#4,#5]{
\begin{figure*}
\vspace*{#5}\
\begin{center}
\begin{minipage}[c]{3in}{
\includegraphics[height=3in]{#1} 
}\end{minipage}\hspace{0.5in}\
\begin{minipage}[c]{3in}{
\includegraphics[height=3in]{#2} 
}\end{minipage}\vspace*{-0.10in} \caption[]{#3}\label{#4}
\end{center}
\vspace*{-0.4in}\
\end{figure*}
}

\def\cxfigure[#1,#2,#3]{
\begin{figure}
\vspace*{4mm}
\begin{center}
 
\epsfxsize=2.5in\
\epsfbox{#1}\
 
\vspace*{-0.10in}\caption[]{#2
} \label{#3}
 
\vspace*{-5mm}
\end{center}
\vspace*{-2mm}
\end{figure}}

\usepackage{xcolor}

\definecolor{cyanish}{rgb}{0,0.8,1.0}
\definecolor{orange}{rgb}{1.0,0.5,0.0}
\definecolor{pink}{rgb}{1.0,0.47,0.6}
\definecolor{light-gray}{gray}{0.95}
\definecolor{jiancolor}{RGB}{0,153,153}
\definecolor{mygreen}{RGB}{50,200,50}
\definecolor{pink}{rgb}{1.0,0.47,0.6}

\definecolor{commentgreen}{rgb}{0.0,0.5,0.0}

\newcommand{\boldparagraph}[1]{\vspace*{0.5ex}\noindent\textbf{#1}\hspace{1em}}

\newcommand{\ignore}[1]{}

\newcommand{\reffig}[1]{Figure~\ref{#1}}
\newcommand{\refsec}[1]{Section~\ref{#1}}
\newcommand{\reftab}[1]{Table~\ref{#1}}
\newcommand{\reflns}[2]{Lines~\hyperref[#1]{\ref*{#1}-\ref*{#2}}}
\newcommand{\us}{\textmu{}s}
\newcommand{\x}[1]{$\times$}

\newcommand{\smahar}[1]{{{\textcolor{cyan}{\textbf{\sffamily $\blacktriangleright$ Suyash: #1 $\blacktriangleleft$}}}}}

\newif\ifcutforspace
\long\def\cutforspace#1{
\ifcutforspace%
        \begingroup%
        \dimen0=\columnwidth
        \advance\dimen0 by -1in%
        \setbox0=\hbox{\parbox[b]{\dimen0}{\protect{\em Cut For Space} #1}}
        \dimen1=\ht0\advance\dimen1 by 2pt%
        \dimen2=\dp0\advance\dimen2 by 2pt%
        \vskip 0.25pt%
        \hbox to \columnwidth{%
                \vrule height\dimen1 width 3pt depth\dimen2%
                \hss\copy0\hss%
                \vrule height\dimen1 width 3pt depth\dimen2%
        }%
        \endgroup%
\fi}

\newcommand{\nova}[1]{NOVA} 
\newcommand{\extfs}[1]{\texttt{Ext4#1}}
\newcommand{\csym}[1]{\texttt{#1}}

\newcommand{\cfunc}[1]{\mbox{\csym{#1}\hspace{-0.1em}\csym{()}}}

\newcommand{\malloc}{\cfunc{malloc}}
\newcommand{\free}{\cfunc{free}}
\newcommand{\mmap}{\cfunc{mmap}}
\newcommand{\fsync}{\cfunc{fsync}}
\newcommand{\msync}{\cfunc{msync}}
\newcommand{\memcpy}{\cfunc{memcpy}}
\newcommand{\memset}{\cfunc{memset}}
\newcommand{\clwb}{\csym{clwb}}
\newcommand{\sfence}{\csym{sfence}}

\newcommand{\eg}{e.g.}

\newif\ifarxiv
\arxivtrue

\newcommand{\papertitle}{Snapshot: Fast, Userspace Crash Consistency for CXL and PM Using msync}

\newcommand{\delete}[1]{{\color{red}}}
\newcommand{\add}[1]{{#1}}

\newcommand{\centerr}[1]{\multirow{2}*{#1}}
\newcommand{\xname}{Snapshot\xspace}
\newcommand{\capitalx}{X}
\newcommand{\PM}{PM\xspace}

\newcommand{\snapshot}{\msync{}\xspace}
\newcommand{\libsnapshot}{\texttt{libsnapshot}\xspace}
\newcommand{\placeholder}[1]{\colorbox{pink}{\sffamily\textcolor{white}{\footnotesize\textbf{#1}}}}

\newcommand{\famus}{FAMS\xspace}
\newcommand{\load}{\textsc{load}\xspace}
\newcommand{\store}{\textsc{store}\xspace}

\newcommand{\cmark}{\color{ForestGreen}{\ding{52}}}
\newcommand{\crossmark}{\color{Maroon}{\ding{56}}}

\newcommand*\circledSolid[1]{\tikz[baseline=(char.base)]{
    \node[shape=circle,fill,inner sep=0.5pt] (char) {\textcolor{white}{#1}};}}
\newcommand{\snap}{\texttt{famus\_snap}}
\newcommand{\fixspacing}{}

\definecolor{burntorange}{rgb}{0.8, 0.33, 0.0}

\ifarxiv
        \newcommand{\cutcamera}[1]{#1}
        \newcommand{\addcamera}[1]{}
\else           
\newcommand{\cutcamera}[1]{}
\newcommand{\addcamera}[1]{{#1}}
\fi

\newcommand{\kyotomax}{8.0$\times$\xspace}
\newcommand{\kyotomin}{1.4$\times$\xspace}
\newcommand{\kvstoreminpmem}{1.2$\times$\xspace}
\newcommand{\kvstoremaxpmem}{2.2$\times$\xspace}
\newcommand{\kvstoremaxmss}{10.9$\times$\xspace}
\newcommand{\btreereadpmdkspeeduppmem}{4.1$\times$\xspace}
\newcommand{\btreeinsertmsyncspeeduppmem}{2.8$\times$\xspace}
\newcommand{\btreeinsertmsynchpspeeduppmem}{463.8$\times$\xspace}
\newcommand{\btreeinsdelsnapshotmsyncminspeedup}{2.8$\times$\xspace}

\ifarxiv
\else
\fi

\usepackage{subcaption}
\usepackage{enumitem}
\usepackage{microtype}
\usepackage{xcolor}
\usepackage{xspace}
\usepackage{tikz}
\usepackage{listings}
\usepackage{pifont} %
\usepackage{pbox}
\usepackage{multirow, tabularx, makecell}

\ifarxiv
\else
  \DeclareFloatSeparators{myfill}{\hskip.013\textwidth plus1fill}
\fi

\ifarxiv
\else
  \makeatletter
  \patchcmd{\@maketitle}
  {\addvspace{0.5\baselineskip}\egroup}
  {\addvspace{-1.2\baselineskip}\egroup}
  {}
  {}
  \makeatother
\fi

\begin{document}

\title{\papertitle}

\date{}

\maketitle
\pagenumbering{arabic} %
\thispagestyle{plain}
\pagestyle{plain}

\begin{abstract}
  Crash consistency using persistent memory programming libraries requires
  programmers to use complex transactions and manual annotations. In contrast,
  the failure-atomic \msync{} (\famus) interface is much simpler as it
  transparently tracks updates and guarantees that modified data is atomically
  durable on a call to the failure-atomic variant of \msync{}. However, \famus{}
  suffers from several drawbacks, like the overhead of \msync{} 
  and the write amplification from page-level dirty data tracking.

  To address these drawbacks while preserving the advantages of \famus{}, we
  propose \xname{}, an efficient userspace implementation of \famus{}.
  \ignore{Like \famus, we argue that achieving crash-consistency should be as
    simple as calling \msync{}. To accomplish this, we propose \xname{}, a novel
    and efficient userspace implementation of the failure atomic \msync{}.}
  \xname{} uses compiler-based annotation to transparently track updates in
  userspace and syncs them with the backing \ignore{persistent memory}
  byte-addressable storage copy on a call to \snapshot{}. By keeping a copy of
  application data in DRAM, \xname improves access latency. Moreover, with
  automatic tracking and syncing changes only on a call to \snapshot{}, \xname
  provides crash-consistency guarantees, unlike the POSIX \msync{} system call.

  \cutcamera{For a KV-Store backed by Intel Optane running the YCSB benchmark, \xname
  achieves at least \kvstoreminpmem speedup over PMDK while significantly
  outperforming conventional (non-crash-consistent) \msync{}.} On an emulated CXL
  memory semantic SSD, \xname outperforms PMDK by up to 10.9$\times$ on all but
  one YCSB workload, where PMDK is 1.2$\times$ faster than \xname. Further,
  Kyoto Cabinet commits perform up to \kyotomax faster with \xname than its
  built-in, \msync{}-based crash-consistency mechanism.
\end{abstract}

\IEEEpeerreviewmaketitle

\begin{table}[b]
  \begin{spacing}{1}
    \justifying 
    {
      \justifying\setlength\parindent{0pt}
      \footnotesize
      \ifarxiv
        \vspace{0.5cm}
        \noindent
        \rule{1cm}{0.4pt}\\
        {
          \begin{spacing}{1}
            \noindent{}A shorter version of this paper appeared in the proceedings of
            \textit{ICCD'23}, citation: \justifying

            Mahar, Suyash, et
            al. ``Snapshot: Fast, Userspace Crash Consistency on CXL Using msync.'' 2023
            IEEE 41st International Conference on Computer Design (ICCD). IEEE,
            2023.
          \end{spacing}
        }
      \fi 
    }
  \end{spacing}
\end{table}

\ifarxiv\else\begin{spacing}{0.98}\fi
\section{Introduction}
\label{sec:introduction}

Recent memory technologies like CXL-based memory semantic
SSDs~\cite{samsung-ssd}, NV-DIMMs~\cite{nv-dimm}, Intel Optane
DC-PMM\ignore{~\cite{optane}}, and embedded non-volatile memories~\cite{reram-soc} have
enabled byte-addressable, non-volatile storage devices.  However, achieving crash
consistency on these memory technologies often requires complex programming
interfaces.  Programmers must atomically update persistent data using
failure-atomic transactions and carefully annotated \load{} and \store{}
operations, significantly increasing programming
complexity~\cite{liu2020cross}.

The \msync{} system call offers a simpler interface for durability.  The
programmer maps a file from the persistent media into the virtual memory and
calls \msync{} to make any changes durable.

The \msync{} interface, however, makes no crash-consistency guarantees. The OS
is free to evict dirty pages from the page cache before the application calls
\msync{}. A common workaround to this problem is to implement a
write-ahead-log~\cite{jhingran1992analysis, postgresql, kyotocabinet} (WAL),
which \ignore{acts as an undo or redo log for the updates, allowing} allows
recovery from an inconsistent state after a failure. However, crash consistency
with WAL requires an application to call \msync{}/\fsync{} multiple times to
ensure application data is always recoverable to a consistent state after a
crash.

Park et al.~\cite{failureatomicmsync} overcome this limitation by enabling
failure-atomicity for the \msync{} system call. Their implementation, \famus{}
(failure atomic \msync{}), holds off updates to the backing media until the
application calls \msync{} and then leverages filesystem journaling to apply
them atomically. \famus is implemented within the kernel and relies on the OS to
track dirty data in the page cache.

OS-based implementation, however, suffers from several limitations:

\ignore{Despite the failure-atomicity guarantees of \famus's \msync, it suffers
  from several of the limitations of the POSIX \msync{} that make it unfit for
  persistent memory programming:}

\emph{(a) Write-amplification on \msync{}:} The OS tracks dirty data at page
granularity, requiring a full page writeback even for a single-byte update\cutcamera{,
wasting memory bandwidth on byte-addressable persistent devices}. \ignore{This
  interface works for disk-based backing media where the minimum write size is
  often the same as the page size (4 KiB). However, For byte-addressable
  memories, like the persistent memory, this wastes bandwidth.} \ignore{Further,
  with the increase in total available DRAM approaching tens of terabytes,
  programmers often turn to huge pages to minimize TLB misses, and the resulting
  expensive page table walks.} Using 2~MiB huge pages to reduce TLB pressure
exacerbates this problem.

\ignore{These PM programming interfaces were developed as a low-overhead
  alternative to the POSIX-based \msync{} interface. In the traditional
  filesystem interface, the application modifies the in-DRAM copy of its data
  and calls \msync{} to ensure the data is durable on the non-volatile
  media. This interface simplifies the programming effort
  significantly. However, in-place modified data of a memory-mapped files can
  reach the non-volatile backing media before the \msync{} is called. Park et
  al.~\cite{failureatomicmsync} solve this problem by ensuring memory-mapped
  file's in-memory data is only made durable on a call to \msync{}. Thus
  providing stronger guarantee than the POSIX \msync{}.}

\emph{(b) Dirty page tracking overhead:} \ignore{Since \famus relies on the page
  table to track dirty pages, on every \msync{}, the kernel scans the page table
  to find dirty pages and writes them to the backing media. Moreover, to
  maintain TLB coherency, the kernel must flush the TLB after clearing access
  and dirty bits\ignore{~\cite{amit2017optimizing}}, adding significant overhead
  to every \msync{} call.}  \famus{} relies on the page table to track dirty
pages; thus, every \msync{} requires an expensive page table scan to find dirty
pages to write to the backing media. Moreover, since the OS is responsible for
maintaining TLB coherency, the kernel must perform a TLB flush after clearing
the access and dirty bits\cutcamera{~\cite{amit2017optimizing}}, adding
significant overhead to every \msync{} call.

\emph{(c) Context switch overheads:} Implementing crash consistency in the
kernel (\eg, \famus) adds context switch overhead to every \msync{} call,
compounding the already high overhead of tracking dirty pages in current
implementations. \cutcamera{While this overhead was acceptable for slower devices like
HDDs, modern flash devices with 10s of \textmu{}s of
latency~\cite{lowlatencyflash}, this is a significant bottleneck. }

\ignore{Third, since the operating system manages the page cache, the pages
  might get evicted by the OS before the application calls
  \msync{}. Applications often solve this by implementing a write-ahead-log
  (WAL). The write-ahead log is an undo or redo log that the application writes
  to before modifying the data in place, allowing the application to recover any
  partially durable changes on a crash or failure. Park et
  al.~\cite{failureatomicmsync} solve this problem by ensuring memory-mapped
  file's in-memory data is only made durable on a call to \msync{} using
  filesystem journalling. While their implementation provides a stronger
  guarantee than the POSIX \msync{}, their solution can only track the changes
  at page granularity and still requires a system call. Thus, it still suffers
  from the write-amplification for Persistent Memory.}

In this paper, we address the shortcomings of \famus{} with \xname, a drop-in,
userspace implementation of failure atomic \msync{}. \ignore{\xname uses novel,
  compiler-based store-tracking to implement a failure atomic \msync{} in
  \emph{userspace}.} \xname transparently logs updates to memory-mapped files
using compiler-generated instrumentation, implementing fast, fine-grained crash
consistency.  \xname{} tracks all updates in userspace \ignore{at sub-page
  granularity\ }and does not require switching to the kernel to update the
backing media.

\xname works by logging \store{}s transparently and makes updates durable on the
next call to \msync{}. \ignore{To achieve this, \xname{} combines a compiler
  pass and a runtime library that work together to provide crash consistency.
  \xname{}'s compiler pass annotates all stores that can point to a heap
  location with a call to an undo-logging function before modifying the
  in-memory state.} During runtime, the instrumentation checks whether the store
is to a persistent file\ifarxiv\footnote{We refer to a file stored on any
  byte-addressable storage media as a persistent file.}\fi and logs the data in an
undo log.

\ignore{In \xname, when an application wants to make its data durable, it calls
  \snapshot{}, and \xname copies all modified data from the working memory to
  the backing store. }
\ignore{Unlike traditional filesystems where the data might reach the backing
  store before a call to \msync{}, in \xname{}, the modified data is made
  durable only on a call to \snapshot{}. This allows the application to take
  advantage of crash consistency without manually maintaining a write-ahead
  log.}

\xname's ability to automatically track modified data allows applications to be
crash-consistent without significant programmer effort.  \cutcamera{For example,
  \xname's automatic logging enables crash consistency for volatile data
  structures, like shared-memory allocators, with low-performance overhead.}

\ignore{For example, to allocate objects in a memory-mapped file, applications
  can often use a shared memory allocator instead of writing a custom persistent
  memory allocator needed for TX-based crash-consistency programming libraries.}

\ignore{\xname{} allows porting existing filesystem-\msync{} based applications
  to PM, resulting in a significant speedup. Applications based on \xname
outperform PMDK across all workloads while requiring significantly less
programming effort.}

\xname{} makes the following key contributions:

\textbf{(a) Low overhead dirty data tracking for \msync{}}. \xname{} provides
fast, userspace-based dirty data tracking and avoids write-amplification of the
traditional \msync{}.

\textbf{(b) Accelerating applications on byte-addressable storage devices.}
\xname{} enables porting of existing \msync{}-based crash consistent
applications to persistent, byte-addressable storage devices with little effort
(e.g., disabling WAL-based logging) and achieves significant speedup.

  \ignore{\item \textbf{Easy crash consistency.}  \xname{} reduces programmer
    persistent memory programming effort to build persistent memory programs by
    making volatile programs crash consistent. \xname{}'s ability to convert
    volatile programs relaxes the requirement for using special-purpose, PM
    specific allocators for crash-consistent applications, simplifying
    application development.}

\cutcamera{  \textbf{(c) Implementation space exploration for fast writeback}. We study the
  latency characteristics of NT-stores and \clwb{}s and use the results to tune
  \xname{}'s implementation and achieve better performance. \cutcamera{These
    results are general and can help accelerate other crash-consistent
    applications.}}

\ignore{We compare our work with PMDK and \msync{} configured with two different page
sizes (4 KiB and 2 MiB) across several benchmarks, including linked list, b-tree
and KV-Store. We also evaluate kyoto cabinet compare \xname{} against a
crash-consistent software using \msync{}.}

We compared \xname against PMDK and conventional \msync{} (as \famus{} is not
open-sourced) using Optane DC-PMM and emulated CXL-based memory semantic SSDs
(DRAM backed by flash media over CXL~\cite{samsung-ssd}). \cutcamera{For b-tree
  insert and delete workloads running on Intel Optane DC-PMM, \xname performs as
  well as PMDK and outperforms it on the read workload by
  \btreereadpmdkspeeduppmem. Moreover, \xname outperforms non-crash-consistent
  \msync{} based implementation by \btreeinsertmsyncspeeduppmem with 4~KiB page
  size and \btreeinsertmsynchpspeeduppmem with 2~MiB page size for inserts.}
For KV-Store, \xname outperforms PMDK by up to \kvstoremaxpmem on Intel Optane
and up to \kvstoremaxmss on emulated CXL-based memory semantic SSD.  Finally,
\xname performs as fast as and up to \kyotomax faster than Kyoto Cabinet's
custom crash-consistency implementation.\looseness=-1

\ignore{\xname thus significantly improves performance while reducing the programming
effort by not requiring the programmer to annotate updates or manually wrap them
in transactions.}

\ignore{The rest of the paper is structured as follows: First, we discuss the background
of using \msync{} for persistent memory programming and its limitations
(\refsec{sec:background}). Next, we describe \xname in Sections
\ref{sec:overview}-\ref{sec:implementation}.  In \refsec{sec:results}, we
evaluate \xname using several workloads, followed by a discussion on the related
works in \refsec{sec:related}.}

\section{Background and Motivation}
\label{sec:background}

\ifarxiv

  \cutcamera{To understand how the \msync{}-based programming interface can work
    on persistent devices, the following section presents a brief survey of
    byte-addressable storage devices. This is followed by a discussion of crash
    consistency based on a semantically strengthened \msync{}
    interface. Finally, we discuss an existing implementation of
    crash-consistent \msync{}, \famus, and its limitations.}

\fi

\makeatletter
  \@namedef{figure}{\killfloatstyle\def\@captype{figure}\FR@redefs
    \flrow@setlist{{figure}}%
    \columnwidth\columnwidth\edef\FBB@wd{\the\columnwidth}%
    \FRifFBOX\@@setframe\relax\@@FStrue\@float{figure}}%
\makeatother

{\newfloatcommand{capbtabbox}{table}[][0.49\linewidth]
  \begin{figure}%
    \CenterFloatBoxes%
    \setlength\fboxsep{0pt}\setlength\fboxrule{0.75pt}
    \begin{floatrow}%
      \ffigbox[0.9\linewidth][]{%
        \includegraphics[width=\linewidth]{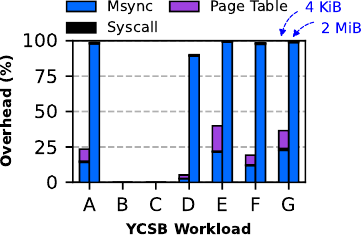}%
      }{\vspace{4pt}\caption{\msync{}-based KV-Store perf. breakdown (4~KiB and 2~MiB%
          \ pages).}\label{fig:msyncoverhead}}%
      \ttabbox[\linewidth][]{%
        \vspace{10pt}
        {\ifarxiv%
\fi%
\centering
  \fontsize{7}{10}
  \selectfont
  \setlength{\tabcolsep}{1pt}
  \renewcommand\arraystretch{0.8}%
    \setlength\extrarowheight{0pt}
  \begin{tabular}{@{\hspace{-2pt}}ccc@{\hspace{-2pt}}}
    \hline\rule{0mm}{2.5mm}
    \textbf{Device}         & \textbf{Interface}       & \textbf{Technology}             \\[0.25mm]\hline\rule{0mm}{2.5mm}
    \centerr{Optane PM}     & \centerr{Mem. Bus}       & {PM \& Internal}               \\
                            &                          & {caches~\cite{wang2023nvleak}} \\[0.25mm]\hline\rule{0mm}{2.5mm}
    Mem. Semantic           & \centerr{CXL} & Flash + Large                  \\ 
    SSDs~\cite{samsung-ssd} &                          & DRAM cache                     \\[0.25mm]\hline\rule{0mm}{2.5mm}
    NV-DIMMs\cutcamera{~\cite{nv-dimm}} & Mem. bus                 & DRAM                           \\[0.25mm]\hline\rule{0mm}{2.5mm}
    Embedded                & {Internal}               & \centerr{ReRAM}                \\
    NVM~\cite{reram-soc}    & {Bus}                    &                                \\
    \hline
  \end{tabular}}%
      }{\caption{Byte-addressable, persistent storage devices.}\label{tab:storage-devs}}%
    \end{floatrow}
  \end{figure}
}

\subsection{Byte-addressable Storage Devices}
Recent advances in memory technology and device architecture have enabled a
variety of storage devices that support byte-addressable persistence. \cutcamera{These
  devices communicate with the host using interfaces like CXL.mem\cutcamera{~\cite{cxl2}},
  DDR-T~\cite{aepperf}, or DDR-4~\cite{nv-dimm} and rely on flash,
  3D-XPoint\ignore{~\cite{3d-xpoint}}, or DRAM as their backing media, as shown in
  \reftab{tab:storage-devs}.}
\ifarxiv{}

\else\fi
These devices share a few common characteristics\addcamera{\
  (\reftab{tab:storage-devs})}: (1) they offer byte-level access to persistent
data, (2) they require special instructions (e.g., cache-line flush) to ensure
persistence, and (3) they are generally slower than DRAM. Later, in
\refsec{sec:overview}, we will explain how \xname takes advantage of these
properties of emerging memories to implement a fast, userspace-based
\snapshot{}.

\subsection{\raggedright{}Filesystem-based Durability and \texttt{msync}}%
\ignore{Programs relying on the filesystem for durability often use the \msync{}
  system call and techniques like write-ahead logging to achieve crash
  consistency.  However, }

The POSIX \msync{} system call guarantees persistency but provides no
atomicity. Part of the dirty data can reach the storage device before the
application calls \msync{}.

To achieve atomicity, applications often use a write-ahead-log
(WAL)~\cite{sqlite-wal,kyotocabinet} to write the modified data to a log and
then change the memory location in place. Applications (\eg, Kyoto
Cabinet~\cite{kyotocabinet}) issue two \msync{} every time they need to
atomically update a mapped file, one to persist the write-ahead-log and the
second to persist the application updates. Once the data is updated and durable
with \msync{}, the application can drop the log.

\ignore{ standard guarantees a
  call to \msync{} with the flag \texttt{MS\_SYNC} would only return when the
  changes in the corresponding region have been durable. Like PM-based
  persistency, the application uses logging to ensure consistency after suffering
  a failure. Applications often use a write-ahead-log (WAL) to write the changes
  to a separate memory region, make it durable using \msync{}, and then make the
  actual changes.}

\ignore{Even though the application data is mapped to DRAM and the backing media
  is not directly accessible, applications still need a log because the
  filesystem might propagate dirty pages to the underlying media.}

Although applications using \msync{} and WAL to achieve crash consistency
directly run off of DRAM (when memory-mapped), they suffer from the overhead of
context switches, page table scanning (for finding dirty pages), and TLB
shootdowns (to clear access/dirty bit for the page table). This overhead is
negligible compared to the access latency of disks and SSDs, but when running on
NVM or memory semantic SSDs, the overhead of performing an \msync{} dominates
the application's runtime. \cutcamera{\reffig{fig:msyncoverhead} shows the \% of runtime
  spent on the \msync{} call, the context switch overhead for the \msync{} call,
  and the TLB shootdown overhead across the YCSB workloads for PMDK's KV-Store,
  modified to use \msync{}. For 2~MiB pages, \msync{}'s overhead is up to 100\% of
  the execution runtime.}\looseness=-1

\cutcamera{With DAX-mapped files, although application \load{}s and \store{}s
  are directly to the storage media (\eg, Optane) and filtered through caches, the
  \msync{} system call still provides no atomic durability guarantee. On \msync on
  a DAX-mapped file, the kernel simply flushes the cachelines of all dirty pages
  to the persistent device.}

Programmers can use a userspace transactional interface (\eg, PMDK) to avoid
performance bottlenecks of the \msync{} system call. While this helps with the
software overhead, PMDK requires programmers to carefully annotate variables,
wrap operations in transactions, and use non-standard pointers. \cutcamera{These
  additional requirements make crash-consistent programming with PMDK
  hard~\cite{corundum} and
  error-prone~\cite{liu2020cross,neal2020agamotto,liu2021pmfuzz,Liu:2019:PAF,gorjiara2021jaaru}.}

  \subsection{Programming with \famus}

  \begin{figure}
    \includegraphics[width=\linewidth]{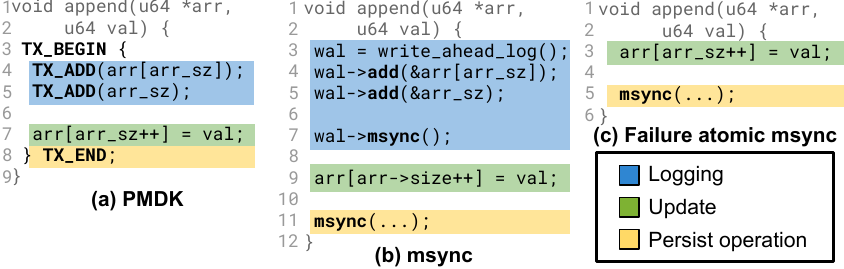}
    \centering
    \caption{Comparison of PM programming techniques using an \texttt{append()}
      function that appends a new entry at the end of a persistent array. \msync{}
      calls in (b) and (c) persist the entire write-containing memory range.}
    \label{fig:cxlbuf-comparison}
  \end{figure}

  \famus fixes shortcomings of the POSIX \msync{} and simplifies crash consistency
  with a failure-atomic \msync{} interface. \famus lets the
  programmer %
  call \msync{} to guarantee that the updates since the last call to \msync{} are
  made atomically durable. \cutcamera{Despite \famus{}'s simpler programming
    model,\ignore{ that results in less code to write and fewer bugs, \famus{}}
    its kernel-based implementation suffers from performance overheads.}

  \reffig{fig:cxlbuf-comparison} compares PMDK, traditional \msync{}-based
  WAL\cutcamera{~\cite{sqlite-wal,kyotocabinet}}, and \famus{} using an \texttt{append()} method
  for an array. Unlike PMDK or \msync{}, \famus does not require
  the programmer to manually log updates\cutcamera{\ either using an undo-log or a write-ahead
    log}.

  In the \famus{} variant (\reffig{fig:cxlbuf-comparison}c), the application maps
  a file into its address space. Next, the application updates the mapped data
  using \load{}s and \store{}s (Line 3) and, finally, calls \msync{} when the data
  is in a consistent state (Line 5). \famus{} ensures that the backing file
  always reflects the most recent successful \msync{} call, which contains a
  consistent state of application data from which the application may recover.
  \ifarxiv

  \else \fi \famus implements failure-atomicity for a file by disabling
  writebacks from the page cache. \cutcamera{When an application calls \msync{}
    on a memory-mapped file, \famus{} uses the JBD2 layer of \extfs{} to journal
    both metadata and data for the file.}  In contrast, PMDK and WAL-based crash
  consistency require programmers to annotate updated memory locations
  manually. Lines 4-5, \reffig{fig:cxlbuf-comparison}a for PMDK, and Lines 3-7,
  \reffig{fig:cxlbuf-comparison}b for WAL-based crash
  consistency. \cutcamera{While PMDK's C++ interface does alleviate some of
    these limitations, it still requires the use of annotations and
    fat-pointers, \eg, \texttt{persistent\_ptr<MyClass> obj}.}\looseness=-1

  Despite \famus's simpler programming interface, applications still suffer from
  kernel-based durability's performance overhead (\eg, context switch overhead,
  page table scanning, etc.).

  \ignore{Next, we will look at \xname and how it offers improved performance over
    current \msync based crash-consistency implementations.}

  \ignore{\subsubsection{Persistent memory allocators}
    Since persistent memory is managed using file
    systems~\cite{ext4dax,NOVA,fortis}, applications allocate objects using
    memory-mapped files. 

    Applications cannot use heap allocators such as
    glibc~\cite{gloger2006ptmalloc} or jemalloc~\cite{evans2006scalable} to allocate
    memory from these files.  Moreover, persistent memory programs require
    special-purpose memory allocators to ensure that the application can recover the
    memory allocator state in the face of a crash and prevent any permanent memory
    leak or memory corruption. Finally, volatile memory allocators work with virtual
    memory pointers, which may be invalid if the file is mapped to a different
    address on restart. Many PM programming libraries include their own
    allocator~\cite{pmdk,corundum,liu2017dudetm,cai2020understanding} that handles
    object allocation and logging. Thus unlike volatile programming, PM programs
    need to use special-purpose allocators that are closely coupled with the logging
    system and the public API.
  }
\section{Overview}
\label{sec:overview}

\xname{} overcomes \famus's limitations by providing a drop-in, userspace
implementation of failure atomic \msync{}, resulting in a significant
performance improvement for crash-consistent applications. To provide low
overhead durability, \xname introduces a compiler-based mechanism to track dirty
data in userspace. \xname records these updates in an undo log that is
transparent to the programmer. On \msync{}, \xname updates the persistent
storage locations recorded in the log\ignore{. (e.g., NVM~\cite{optane} or
  memory semantic SSDs~\cite{samsung-ssd})}.  In case of a failure, \xname can
use the log to roll back any partially durable data.

\ignore{In \xname{}, like \famus{}, the updates propagate atomically to the
  backing media when the programmer calls \msync{}, making updates to the
  persistent data failure-atomic.  \xname instruments stores to track updates to
  the memory-mapped file with sub-page granularity, avoiding expensive system
  calls.}

\ignore{\xname{} tracks updates to the memory-mapped file with sub-page
  granularity using function calls, automatically inserted by its compiler pass
  before every application \store{}. These instrumented function calls log
  updates performed by the application, greatly simplifying persistent memory
  programming while avoiding expensive system calls.}

Since \xname{} is implemented in userspace, it avoids the overhead of switching
to the kernel and managing TLB coherency. Using \xname{}, the application
synchronously modifies data only on the DRAM, speeding up the execution. At the
same time, \xname{} maintains a persistent copy on the backing media and
automatically propagates all changes to the persistent copy on an \msync{}.

\cutcamera{As \xname{} is built on the \msync{} interface, programmers can port any
conventional application written for \msync{}-based crash consistency with
minimal effort to benefit from automatic dirty-data tracking while significantly
improving runtime performance. \xname{}'s userspace implementation enables
legacy disk-based applications to take advantage of faster access times and
direct-access (DAX)\ignore{~\cite{ext4dax}} storage without requiring extensive
application rewrites.}

\section{Implementation}
\label{sec:implementation}
\ifarxiv

  \xname{} is implemented as a combination of its compiler pass and a runtime
  library, \libsnapshot. \xname{}'s compiler instruments every store instruction
  that can write to the heap using a call to an undo-log function.  The runtime
  library, \libsnapshot, provides runtime support for \xname{}: implementing the
  logging function and \xname{}'s \snapshot.

\fi

Next, we will discuss how the programming interface and logging for \xname{} are
implemented, followed by the various optimizations possible in \xname{} to improve
its performance.

\subsection{Logging, Instrumentation, and \snapshot{}}
\xname{} tracks updates to persistent data by instrumenting each \store{} in the
target application with a call to the logging function (instrumentation). During
runtime, the instrumentation takes the \store{}'s address as its argument,
checks if the \store{} is to a persistent file, and logs the destination
memory location to an undo log.

\boldparagraph{Logging and Recovery.}
\xname{}'s undo log lives on persistent media to enable recovery from crashes
during an \msync{} call. When an application calls \snapshot{}, \xname{} reads
the addresses from the undo log entries and copies all modified locations from
DRAM to the backing media. The call to \snapshot{} only returns when all the
modified locations are durable. If the system crashes while copying the
persistent data, on restart, \xname{} uses its undo-log to undo any changes that
might have partially persisted.

While \xname{} maintains per-thread logs, calls to \msync{} persist data from
all threads that have modified data in the memory range. \xname{} maintains a
thread-local log to keep track of modified locations and their original values.
\xname{} provides limited crash-consistency guarantees for multithreading,
similar to PM transactional libraries like PMDK. For example, PMDK prohibits
programmers from modifying shared data from two threads in a single transaction.
Similarly, in \xname{}, the program should not modify a shared location from two
threads between two consecutive \msync{}s.

\ignore{\nfigure[log-format.pdf,\xname{}'s thread-local log for tracking changes
between \msync{}s and for recovery on a crash.\fixspacing,fig:log-format]}

\boldparagraph{Log Format.}  Logs in \xname{} are per-thread and store only the
minimal amount of information needed to undo an interrupted \msync{}.
\ignore{\reffig{fig:log-format} shows the log layout that \xname{} uses to track
  in-memory modifications. Each log holds its} Logs hold their current
state\ignore{(\texttt{State})}, that is, whether a log holds a valid value.
\cutcamera{The log also maintains a tail \ignore{(\texttt{Tail})} that points to
  the next free log entry and the size of the log for use during recovery.  Each
  log entry in the log is of variable length. The log entry consists of the
  address\ignore{ (\texttt{Addr})}, its size\ignore{ (\texttt{Size})} in bytes,
  and the original value at the address\ignore{ (\texttt{Content})}.}

While \xname tracks all store operations to the memory-mapped region, POSIX
calls such as \memcpy{}, and \texttt{memmove()} are not instrumented as they are
part of the OS-provided libraries. To solve this, \libsnapshot{} wraps the calls
to \memcpy, \texttt{memmove()}, and \memset{} to log them directly and then
calls the corresponding OS-provided function. \cutcamera{While \xname{} catches
  some of these functions, applications relying on other functions, e.g.,
  \texttt{strtok()}, would need to recompile standard libraries (glibc, muslc,
  etc.) with \xname to be crash-consistent.}\looseness=-1

\boldparagraph{Logging Design Choices.} Despite implementing undo-logging,
\xname{} only needs two fences per \snapshot to be crash-consistent, as it does
not need to wait for the undo-logs to persist before modifying the DRAM
copy. This contrasts with PMDK (which also implements undo-logging), where every
log operation needs a corresponding fence to ensure the location is logged
before modifying it in place. Redo-logging persistent memory libraries eliminate
this limitation and only need two fences per transaction. Redo logging, however,
requires the programmer to interpose both the loads and stores to redirect them
to the log during a transaction, resulting in higher runtime overhead. \xname{},
on the other hand, only interposes store instructions which always write only to
the DRAM and avoids any redirection.

\subsection{Optimizing \xname{}}
\ifarxiv
  \xname{} includes a range of optimizations to maximize its performance. In
  particular, it must address challenges related to the cost of range tracking and
  reducing the required instrumentation.
\fi

\ifarxiv%
  \subsubsection{Low-cost Range Tracking}
\fi%
Since \xname{}'s compiler has limited information about the destination of a
\store{}, on every call, the instrumentation checks if the logging request is to
a memory-mapped persistent file.  \xname{} simplifies this check by reserving
virtual address ranges for DRAM (\emph{DRAM range}) and the backing memory
(\emph{persistent range}) when the application starts. Reserving ranges on
application start makes checks for write operations a simple range
check. \ignore{\reffig{fig:cxlbuf-address-space} shows how a file from a
  persistent device is mapped to \emph{both} these ranges at the same offset.}
\cutcamera{In our implementation, we reserve 1~TiB of virtual address space for both ranges
to map all persistent-device-backed files. This range is configurable and is
limited only by the kernel's memory layout.
Further, copying a location from DRAM to the backing media now only needs 
simple arithmetic, i.e., copy from offset in the DRAM range
to the same offset in the persistent range.}

\ifarxiv%
  \subsubsection{Fewer Instrumentations}
\fi%
Instrumenting stores indiscriminately results in useless calls for
\texttt{store}s that cannot write to persistent locations (e.g., stack
addresses). \xname{} reduces this overhead by tracking all stack allocations in
a function at the LLVM IR level during compilation. Next, \xname{} instruments
only those \texttt{store}s that may not alias with any stack-allocated
addresses, reducing the amount of unnecessary instrumentation.

\ignore{\subsubsection{Compressed log entries}
For certain updates to the memory mapped data, \xname{} can compress the resulting
log entry to take significantly less space, improving the performance of the
logging operation. This is implemented for the \memset{} operation, where on
intercepting the library call, \xname{} creates a single log entry with the
parameters to the}

\ignore{\subsubsection{Disabling Instrumentation for Functions}
If a function performs no persistent media access, \xname{} allows the programmer
to skip the function during instrumentation using function
annotations. \ignore{This allows \xname{} to skip any undo log calls that are not
  to persistent memory, but cannot be resolved at compile time.} However, we do
not use this annotation in our evaluation and let the compiler instrument all
possible heap stores.
}

\cutcamera{

  \subsection{Optimizing Backing Memory Accesses}
Flush and fence instructions needed to ensure crash consistency add significant
runtime overhead. To understand and reduce this overhead, we study the relative
latency of write+\clwb{} vs. NT-Store instructions and find that non-temporal
stores, particularly those that align with the bus's transfer size, result in
considerable performance improvement over the \clwb{} instruction.

\delete{behavior of the \clwb{}, \sfence{}, and NT-Store instructions by measuring how
\sfence{} adds overhead to \clwb{} and NT-Store instructions and by measuring
the relative latency of write+\clwb{} vs. NT-Store instructions. Further, based
on these observations, \xname includes optimization to align writes to make them
compatible with NT-stores.}

While we perform \delete{these} the experiments on Intel Optane DC-PMM, we
expect the results and methodology to be similar on other storage devices as
they have similar memory hierarchies (volatile caches backed by byte-addressable
storage devices).

\begin{figure}
  \centering
  \includegraphics[width=0.9\linewidth]{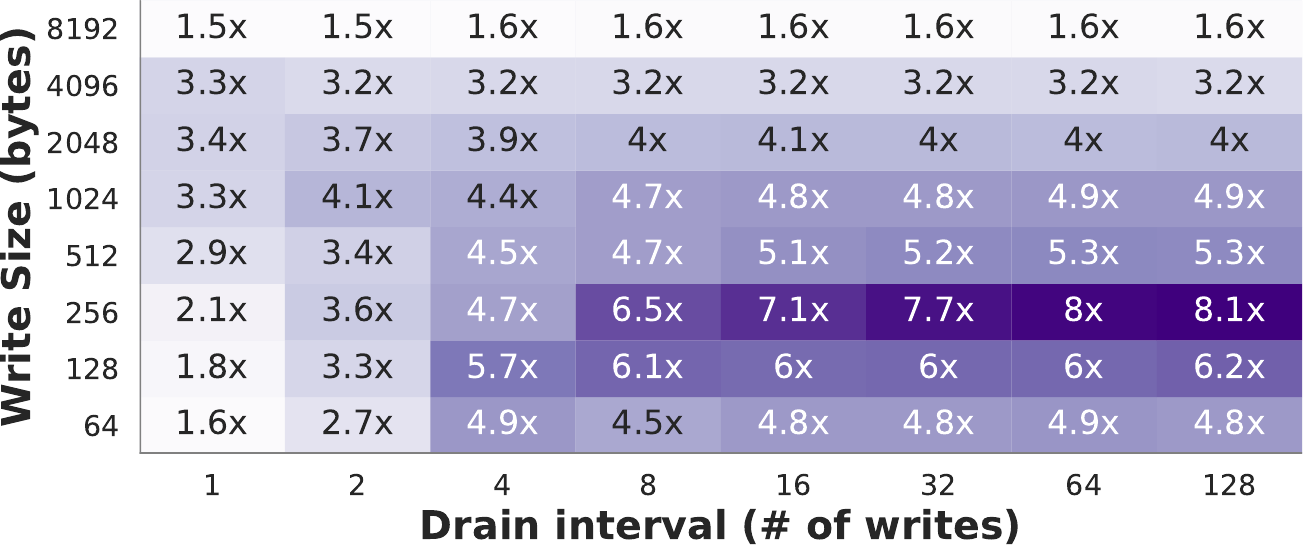}
  \caption{Speedup of NT-stores over \clwb{} instructions for PM
    writes. NT-stores always outperform write+\clwb.}
  \label{fig:clwbvsntstore}
\end{figure}

\delete{From our experiments, we answer two important questions regarding behaviors of
the \sfence{} and \clwb{} instructions:

\paragraph{When to fence after a write?}{We observe that increasing
  flush-to-fence distance improves performance by relaxing the store ordering
  constraints.}

\paragraph{How to choose between non-temporal stores and flushes?}
Non-temporal stores, particularly those which align with the bus's transfer
size, result in considerable performance improvement over the \clwb{}
instruction.

Next, we will explore the answers to these questions in detail and use the
findings to optimize \xname{}.}

\delete{\boldparagraph{When to fence after a write?} 
\xname{} optimizes its implementation by delaying a fence for as long as
possible after flushing data to the backing memory. We observe that the farther
the fence is from a flush, the smaller its overhead is.

While \xname{} reduces the number of fences, it still has to ensure all logs are
persistent before committing the changes (that is, before returning from an
\snapshot call).

\nfigure[clwbsfencedist.pdf,{Change in latency of store instructions
  (\clwb{} and NT-stores) with increasing write-to-fence distance. The vertical
  axis starts at 1.1~\us{}.}\fixspacing,fig:clwbsfencedist]

To understand the performance benefits of delaying a fence after flushing
persistent data, we use a microbenchmark that flushes a location to the backing
memory and performs a variable number of unrelated operations between the flush
and the fence. \reffig{fig:clwbsfencedist} shows results for three variants of
the microbenchmark:

\begin{enumerate}[leftmargin=15pt, rightmargin=0cm,itemsep=0pt]
\item \textbf{\clwb{} + delayed fence.} The microbenchmark issues a
  \texttt{store}, followed by a flush (\clwb) but delays the fence (\sfence)
  into the future.
\item \textbf{Delayed \clwb{} + fence.} The microbenchmark issues a
  \texttt{store} and delays both the flush (\clwb) and the fence (\sfence)
  operations.
\item \textbf{Non-Temporal (NT) Store + delayed fence.} The microbenchmark
  writes to the memory using an NT-Store and delays the corresponding fence
  (\sfence).
\end{enumerate}

Results show that the farther away a fence operation is from a write/flush, the
smaller the write operation's effective latency is.  We use this observation in
\xname{} and issue a fence as late as possible after a flush (\clwb{}) when
NT-Stores cannot be used. We thus issue a fence for the logging operations only
when the application calls \msync{}.

Issuing fewer fences during the execution reduces the runtime overhead of the
flush instructions by relaxing the ordering constraints.\\
}
 
\reffig{fig:clwbvsntstore} shows the latency improvement from using NT-Stores to
update PM data vs. using writes followed by \clwb{}s. The heatmap measures the
latency of the operation while varying both the write size, that is, the amount
of data written, and the frequency of the fence operation
(\sfence{}). \delete{Unlike \reffig{fig:clwbsfencedist}, where the
  microbenchmark performs unrelated operations to delay a fence, in
  \reffig{fig:clwbvsntstore}, the microbenchmark only generates NT-Store or
  \texttt{store}+\clwb{} traffic.}

From \reffig{fig:clwbvsntstore}, we observe that NT-Stores consistently
outperform \texttt{store}+\clwb{}. Moreover, the performance gap between
\clwb{}s and NT-Stores increases as the fence interval increases. In contrast,
when the write size is increased, the performance only increases until the write
size matches the DDR-T transaction size (256
B)\ignore{~\cite{NVMMEmpirical}}. Since the CXL protocol uses 64B packet size
for v1-2 and 256B for CXL v3, we expect to see maxima for those sizes for
CXL-based byte-addressable storage devices.

Based on this observation, \xname{} always uses NT-Store instructions to copy
modified cachelines from the DRAM copy to the persistent copy on a call to \snapshot{}.\\

}

\ignore{\boldparagraph{Aligning Unaligned NT-Stores.}\label{sec:unaligned-nt-stores}
Even though \reffig{fig:clwbvsntstore} NT-stores have better write performance
(for infrequently accessed data), on x86-64, NT-stores only support writing to a
location aligned to their size (e.g., aligned to 512 bytes for 512-byte
NT-Store). This limitation of NT stores holds back \xname{}'s performance when
writing to unaligned locations.

On a call to \snapshot{}, \xname{} first tries to copy the modified location from
DRAM to the backing media using NT stores if the writes are larger than 4 bytes
and are aligned to the largest possible NT store. However, if a write is not
aligned, \xname{} increases the write size by aligning the start of the write to
the closest NT store for the operation.}

\delete{

  \begin{figure}
    \begin{center}
      \includegraphics[width=0.8\columnwidth]{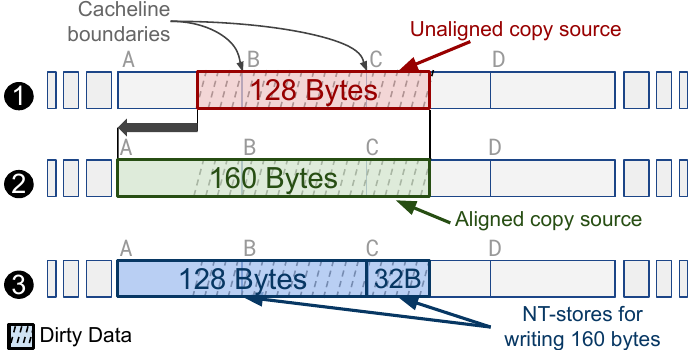} 
      \caption[]{\xname{} aligns the copy start and end to NT-store boundaries, thus
        replacing expensive \clwb{}+\sfence{} with NT-stores for uncached data.}
      \label{fig:unaligned-nt-stores}
    \end{center}
  \end{figure}

  The example in \reffig{fig:unaligned-nt-stores} shows a 128 bytes write
  operation that is unaligned \circledSolid{1}. Since the largest NT-store \xname{}
  can perform is of 128-bytes, it aligns the start of the store to the previous
  128 alignment boundary (start of the cacheline A, \circledSolid{2}). Once
  aligned, \xname{} will issue one 128-bytes NT store to write the first two
  cachelines (cachelines A and B), followed by a 32-byte NT store to write the
  last half cacheline (cacheline C, \circledSolid{3}).\\
}

\ignore{\paragraph{Limitations}
Aligning stores by increasing their size can result in scenarios where the
resulting persistent data is inconsistent. Consider an example where the
programmer writes to an array of structs that are large enough to be persisted
using NT-Stores but are not aligned to any of the instruction sizes. In this
case, \xname{} would increase the write size to align them to the NT-store
boundary, but in doing so, \xname{} will persist more data than needed and might
persist a neighboring struct that is not in a consistent state.

\xname{} allows the programmer to disable this optimization if multiple threads
update adjacent unaligned data. However, since threads often work on isolated
data, unaligned stores can be used without violating
crash consistency. Moreover, in the evaluation, we will measure the performance
improvement using a KV-store with the optimization enabled and disabled
(\refsec{sec:kv-store-results}).}

\cutcamera{

  \boldparagraph{Reducing Backing Memory Accesses}\label{sec:reducing-pm-access}
  To find modified locations to write to the persistent copy from the DRAM copy,
  \xname{} iterates over its undo log. As this log is stored on the backing
  memory, it can be slow to access. This is a result of the log design where
  each log entry is of variable length, thus, accesses to sequential log entries
  result in variable strides and poor cache performance.  As a result, \xname{}
  has to waste CPU-cycles traversing the entries.  To reduce read traffic to the
  backing memory and mitigate this additional overhead, \xname{} keeps an
  additional, in-DRAM list of the updated addresses and their sizes, thus
  avoiding accessing the backing memory.

  While it is possible to split the log to separate log entry's data into a
  different list, log entries would then require more instructions to flush
  them, adding overhead to the critical path.

  \ignore{\xname{} reduces read traffic to the backing memory during an
    \snapshot by keeping an in-DRAM list of the updated addresses. In \xname,
    the log is laid out in the memory as a sequence of log entries
    (\reffig{fig:log-format}). This is done to keep the log layout
    simple. However, the log's layout and the use of persistent media result in
    low cache locality and high cache miss latency when \xname needs to read the
    list of updated addresses on a call to \msync{}. Thus, we keep an additional
    list in the DRAM containing the address and size of the updates. On a call
    to \snapshot, \xname only reads the volatile list and uses it to copy all
    the updated locations from DRAM to the persistent device.}
}

\subsection{Memory Allocator}
While \xname provides a failure atomic \msync{}, applications need to
  allocate and manage memory in a memory-mapped file. 
  \ifarxiv
    Shared memory allocators,
    like \texttt{boost.interprocess}~\cite{boost.interprocess}, provide an easy way to manage memory in a
    memory-mapped file by providing \malloc{} and \free{}-like API. These
    operations, while enable memory management, are not crash-consistent.
  \fi

\cutcamera{However, \xname's ability to automatically log all updates to the
  persistent memory, enables applications to use volatile shared memory
  allocators for allocating memory in a crash-consistent manner.}

To demonstrate \xname's \ifarxiv utility\fi, we use \xname to enable
\texttt{boost.interprocess} to function as a
persistent memory allocator. \texttt{boost.interprocess} allocates objects from
a memory-mapped file, provides API to access the root object as a pointer, and
allocate/free objects while \xname tracks updates and makes changes atomically
durable on an \msync{}.

\cutcamera{\boldparagraph{Decoupling Memory Allocator and Logging} Unlike traditional PM
programming libraries that couple memory allocators and logging techniques,
\xname{} permits any combination of a logging technique and a memory allocator.
For example, PMDK's allocator only supports redo logging. Restricting the
programmer to the specific characteristics of their implementation.  On the
other hand, \xname provides programmer the ability to independently choose the
memory allocator and logging technique, suiting the specific needs of the
workload.

\ignore{Traditional persistent memory programming libraries often require the programmer
to use a specific combination of memory allocator and logging techniques. For
example, PMDK only supports undo-logging application data and its own allocator.

\xname{} presents a new crash-consistency programming architecture by decoupling
the memory allocator from its logging. Since the compiler automatically logs
updates from the memory allocator, \xname{} can independently support a variety
of logging techniques (undo, redo, and hybrid) while also letting the programmer
choose a memory allocator of their choice.}

}

\ignore{This is also true for other programming libraries such as
  Corundum~\cite{corundum}, NV-Heaps~\cite{nvtm}, any others \placeholder{cite
    me}~\cite{citeme}.}

\cutcamera{\subsection{Putting it all Together}
\nfigure[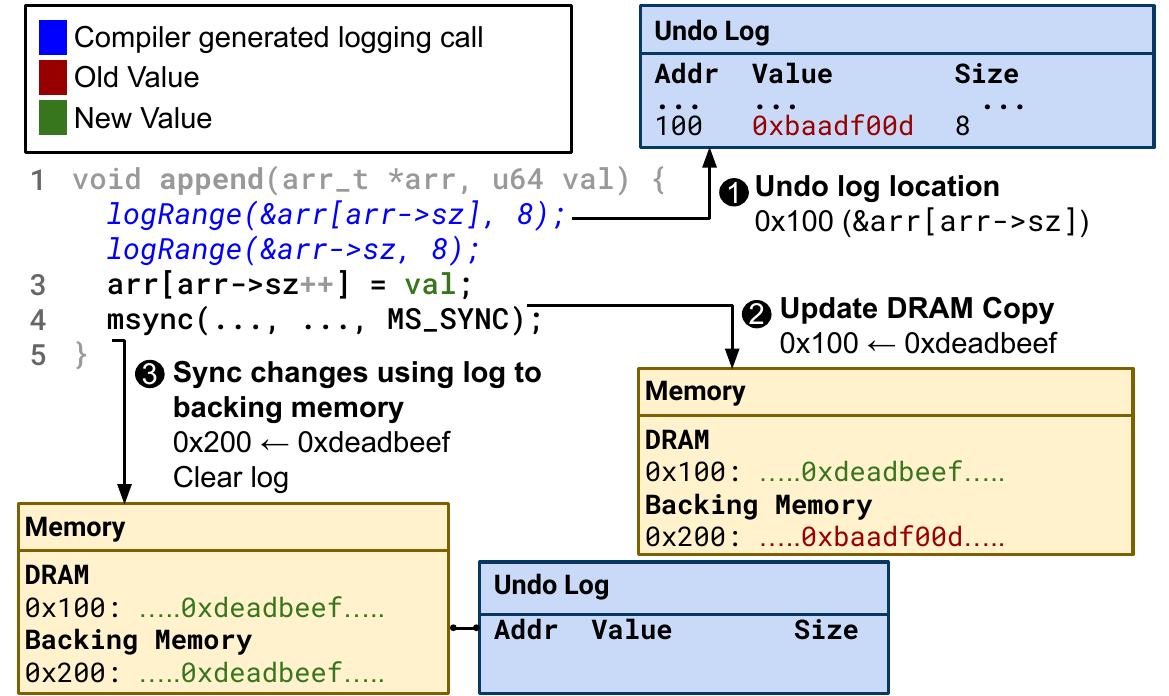,\xname{}'s working. Instrumented binary
calls \texttt{libsnapshot.so}'s logging function for every store. Changes are
atomically durable on \msync{}.\fixspacing,fig:cxlbuf-working]

To see how snapshot works in practice, consider an array \texttt{append()}
function that takes a value and inserts it into the next available slot in the
array. \reffig{fig:cxlbuf-working} shows the implementation of this function
along with the memory and log states as the program executes. In the example,
the instrumented program automatically undo-logs the updated location (i.e.,
call to \texttt{logRange}). For brevity, we only show updates to the array
element, not the array size.

When the program starts executing \circledSolid{1}, the instrumentation calls
the logging function with the address of updated locations
(\texttt{\&arr[arr->sz]} and \texttt{\&arr->sz}), and update sizes.  The
function logs the address by creating a new entry in the thread-local undo log.

Next, \circledSolid{2} the program continues and updates memory locations. Since
the program directly interacts only with the DRAM, the value in the DRAM is
updated, but the value in the backing memory (e.g., \PM) is unchanged.  Finally,
\circledSolid{3} the application calls \snapshot{} to update the backing
memory. On this call, \xname{} iterates over the log to find all locations that
have been updated and uses them to copy updates from DRAM to PM. After updating
\PM with the values from DRAM, \xname{} drops the log by marking it as
invalid. Once \snapshot returns, any failure would reflect the persistent state
of the most recent \msync{}.
}

\cutcamera{\subsection{Correctness Check}
We test our compiler pass and resulting binary to ensure correctness and crash
consistency. We test for crash consistency bugs by injecting a crash into the
program before it commits a transaction when \xname{} has copied all the changes
to the backing store but has not invalidated the log. On a restart, \xname{} should
recover and let the application continue its normal execution. This is only
possible if the compiler pass correctly annotates all the store instructions and
the logging function logs them.
We ran these tests for multiple configurations and inputs and found that the
compiler pass and the runtime recovered the application each time.}

\section{Results}
\label{sec:results}

\ifarxiv
\thisfloatsetup{floatrowsep=none}
\begin{table*}%
  \CenterFloatBoxes%
  \fontsize{7}{10}
  \selectfont
  \begin{floatrow}[2]
    \setlength\fboxsep{0pt}\setlength\fboxrule{0.75pt}
    \ttabbox[\FBwidth]%
    {{
\ifarxiv
\else
  \renewcommand\arraystretch{0.8}%
\fi
\setlength\extrarowheight{0pt}
\setlength{\tabcolsep}{2pt}
\begin{tabular}{|l|l|c|c|c|} 
  \hline\rule{0mm}{2.5mm}
  \centerr{\textbf{Config}} & \centerr{\textbf{Description}}                                            & \textbf{Dirty data} & \textbf{Crash}      & \textbf{Working} \\
                            & \textbf{}                                                                 & \textbf{tracking}   & \textbf{consistent} & \textbf{memory}  \\\hline\rule{0mm}{2.3mm}
  PMDK                      & Intel's PMDK-based implementation.                                        & Programmer (byte)   & \cmark              & PM               \\[0.25mm]\hline\rule{0mm}{2.3mm} 
  \xname{}-NV               & \xname, tracking using undo-log.                                          & Auto., (byte)       & \cmark              & DRAM             \\[0.25mm]\hline\rule{0mm}{2.3mm}
  {\xname{}}                & \xname, tracking using a volatile list (\refsec{sec:reducing-pm-access}). & {Auto., (byte)}     & {\cmark}            & {DRAM}           \\[0.25mm]\hline\rule{0mm}{2.3mm}
  \msync{} 4 KiB            & Page cache mapped, 4 KiB pages.                                           & Auto., OS (4KiB)    & \crossmark          & DRAM             \\[0.25mm]\hline\rule{0mm}{2.3mm}
  \msync{} 2 MiB            & Page cache mapped, 2 MiB pages.                                           & Auto., OS (2MiB)    & \crossmark          & DRAM             \\[0.25mm]\hline\rule{0mm}{2.3mm}
  {\msync{} data journal}   & Page cache, ext4 (\texttt{data=journal}), 4 KiB Pages                     & {Auto., OS (4 KiB)} & {\crossmark}        & {DRAM}           \\[0.25mm]\hline %
\end{tabular}

}}{
      \caption{Evaluated configurations.}
      \label{tab:configs}
    }%
    \ttabbox[\FBwidth]%
    {{

  \renewcommand\arraystretch{0.8}%
  \setlength{\tabcolsep}{2pt}
  \setlength\extrarowheight{0pt}
  \begin{tabular}{|l|l|}    
    \hline\rule{0mm}{2.3mm}
    {CPU}                   & 2 $\times$ Intel 6230, 40 HW threads,          \\[0.25mm]\hline\rule{0mm}{2.3mm}
    DRAM                    & 192~GiB (DDR4)                                 \\[0.25mm]\hline\rule{0mm}{2.3mm}
    \multirow{2}{*}{Optane} & {100 series, 128\x{}12 = 1.5~TiB,}             \\
                            & {AppDirect Mode}                               \\[0.25mm]\hline\rule{0mm}{2.3mm}
    OS \& Kernel            & Ubuntu 20.04.3 \& Linux 6.0.0                  \\[0.25mm]\hline\rule{0mm}{2.3mm}
    Build system            & LLVM/Clang 13.0.1                              \\[0.25mm]\hline\rule{0mm}{2.3mm}
    Block Device            & \centerr{Intel Optane SSD DC P4800\capitalx{}} \\
    (for emulation)         &                                                \\[0.25mm]\hline %
  \end{tabular}
}}{
      \caption{System configuration}
      \label{tab:sysconfig}
    }
  \end{floatrow}
\end{table*}
\fi

\ifarxiv

  To understand \xname's performance, instrumentation overhead, and the impact
  of various optimizations, we evaluate several microbenchmarks, persistent
  memory applications, including Kyoto Cabinet, and crash-consistency
  solutions. Further, to get an estimate of \xname's performance on future
  hardware, we evaluate \xname{} against PMDK on a CXL-based emulated
  memory-semantic SSD.

\fi

\subsection{Failure Atomic \msync{} Implementations}
Three implementations of failure atomic \msync{} are possible candidates for
comparison with \xname{}. The original implementation, \famus, is by Park et
al.~\cite{failureatomicmsync}. The other implementations,
\snap{}~\cite{kelly2019good} and AdvFS~\cite{verma2015failure}
use reflinks and file cloning, respectively, to create shallow copies of the
backing file on \msync{}.

\famus by Park et al.~\cite{failureatomicmsync} is not open-sourced%
\ifarxiv. We \fi
use POSIX \msync{} with data journalling enabled to approximate its
performance. \cutcamera{\famus works by reconfiguring the \extfs{}'s data journal to not
write back to the backing media until the application calls \msync{}. Since
\famus uses data journalling to implement failure atomicity, their
implementation performs similarly to \msync{} on \extfs{} mounted with the
option \texttt{data=journal}, as shown by Park et al.~\cite{failureatomicmsync}.
}
\ignore{Thus, to approximate Park et al.'s \famus's performance, we evaluate
  \msync{} on an \extfs{} filesystem mounted with the \texttt{data=journal}
  flag. We refer to this configuration as \msync{} data journal. Since Park et
  al. showed that the \msync{} data journal performs slightly faster than
  \famus, it acts as an upper bound of \famus' performance.}
\ifarxiv

Implementation of failure atomic \msync{} by Verma et al. on
AdvFS~\cite{verma2015failure} is not open-sourced, however, Kelly's
\snap{}~\cite{kelly2019good} is open-sourced and can be evaluated. \snap{} uses
reflinks to create a snapshot of the memory-mapped file on a call to \msync{}.
\fi

\ifarxiv
\snap{}, however, is much slower than the POSIX \msync{}\cutcamera{ due to slow underlying
\texttt{ioctl(FICLONE)} calls}. In our evaluation, we found that \snap{} is
between 4.57$\times{}$ to 338.57$\times$ slower than \msync{} for the first and
500th calls, respectively. \ignore{For an \msync, after every 2
  updates on 256~MiB region, \msync{} takes 29.5~\us{}, while \snap{} takes
  8.5~ms per call.}

\fi
Since \snap{} is orders of magnitude slower than \msync{}, we do not
evaluate it further.\vspace{-0.12cm}

\cutcamera{
\subsection{Configuration}
\reftab{tab:configs} lists the six different configurations we use to compare
the performance of \xname{}. With PMDK, the workloads are implemented using its
software transactional memory implementation. The \xname{}-NV and \xname{}
implementations are similar, with the difference in how they track dirty data in
DRAM. The \xname-NV implementation uses the undo log to flush the dirty data on
a call to \snapshot.  In comparison, the \xname{} implementation uses a
separate, volatile list to flush the dirty data
(\refsec{sec:reducing-pm-access}).  All implementations of \xname{} otherwise
have this optimization enabled.  \reftab{tab:sysconfig} lists the configuration
used for all experiments in the results section.}

\subsection{\vspace{-0.1cm}Evaluating \xname{} on CXL-based Memory Semantic
  SSDs}\label{sec:mem-sem-ssd}

\cutcamera{
\begin{figure}
  \includegraphics[width=\linewidth]{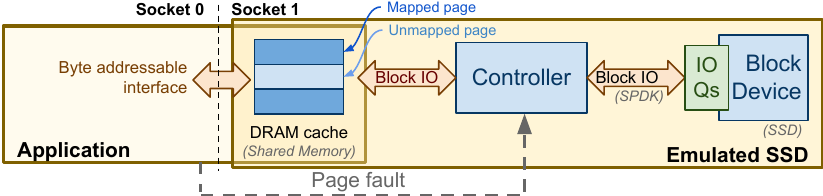}
  {}
  \caption{Emulated memory-semantic SSD architecture.}
  \label{fig:memsemssd}
\end{figure}
}

CXL-attached memory semantic SSDs~\cite{samsung-ssd} are CXL-based devices with
a large DRAM cache backed by a block device. These devices appear as memory
devices to the host processor and support byte-addressable accesses.

To understand how \xname{} would perform on CXL-attached memory semantic SSDs,
we created a NUMA-based evaluation platform.  In our emulation, we implement the
DRAM cache using shared memory and service cache miss from a real SSD in
userspace using SPDK\cutcamera{~\cite{yang2017spdk}}. The target application and
the emulated SSD are pinned to different sockets to emulate a CXL link (similar
to Maruf et al.~\cite{maruf2022tpp}).  \cutcamera{\reffig{fig:memsemssd} shows
  the architecture of our emulated memory semantic SSD.}

In our implementation, the shared memory used to emulate the DRAM cache has only
a limited number of pages (128 MiB) mapped. On access to an unmapped page by the
application, the emulated SSD finds a cold page to evict using Intel
PEBS\cutcamera{~\cite{intelpebs}} and reads and maps the data from the SSD.

Our implementation has a 14.3~\us{} random access latency at a 91.8\% DRAM cache
miss rate and a 2.4~\us{} latency at a 16.3\% DRAM cache miss 
rate. \cutcamera{While these latencies might be high compared to Intel Optane
  DC-PMM, they represent a slower, byte-addressable media and are close to
  low-latency flash, \eg, Samsung Z-NAND~\cite{cheong2018flash}.}

\cutcamera{\subsection{Microbenchmarks}

\ignore{\xname{}'s instrumentation is critical to its performance. }

Next, we use microbenchmarks to study how \xname's store instrumentation affects
performance. \ignore{ and if \xname's implementation limits its scalability in
  multithreading.}

\begin{figure}
  \includegraphics[width=\linewidth]{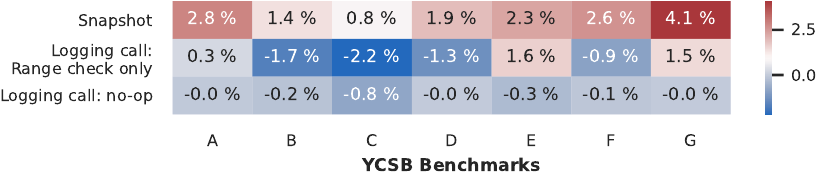}
  {}
  \caption{\xname instrumentation's overhead normalized to no-instrumentation.
    Excludes persistence overhead.}
  \label{fig:instrumentationoverhead}
\end{figure}  

To understand the instrumentation overhead of \xname{}, we run it with and
without instrumentation and logging
enabled. \reffig{fig:instrumentationoverhead} shows the performance of different
variants of \xname{} relative to the ``No Instrumentation'' variant running the YCSB
workload. The ``Logging call: no-op'' variant returns from the logging call
without performing any checks or logging. The ``Logging call: range check only''
measures the execution where the logging call only performs the range check but
does not log any data. Finally, the ``No instrumentation'' variant is compiled
without \xname's compiler pass and thus has no function call overhead. Among
these, only \xname{} logs the modifications and is crash-consistent. In all
other variants, a call to \msync{} is a no-op.

The results show that even with the compiler's limited information about a store
instruction, the overhead from the instrumentation is negligible since stores
are relatively few compared to other instructions.
}

\cutcamera{\boldparagraph{Store Instrumentation
  Statistics.} \label{sec:instrumentation-statistics}
\ignore{\reffig{fig:instrumented-stacked} shows the number of store instructions
  instrumented across three workloads: KV-store, linked list, and b-tree. The
  bar height represents the total number of store instructions in the workload.}
Across the workloads, \xname instrumented 10.8\% of store instructions on
average. Out of all the store instructions, \xname skipped 84.6\% because they
were stack locations, and 4.54\% because they were aliased to stack locations.
In case a location is not a stack location, or aliased to one, \xname errs on
the side of caution and instruments it. During runtime, the instrumentation
checks the store's destination to ensure it is to a persistent location.

\ignore{ the breakdown shows the store instructions skipped because they were
exact matches to stack allocated addresses or if they matched to one of the
stack locations using alias analysis. Since many memory allocations are
function-local and only a few functions modify the heap, instrumented locations
are a small fraction of all the store instructions.

On average, across workloads, 10.8\% of the total number of stores are
instrumented.\\}
}

\ignore{\boldparagraph{Multithreaded Scaling.} \label{sec:multithreadedscaling}
\nfigure[msyncscaling.pdf,{Scaling of \xname{} and \msync{} with
  increasing thread count.},fig:msync-scaling]
To understand the impact of multithreading on performance, we scale the number
of threads and measure the total runtime for a
microbenchmark. \reffig{fig:msync-scaling} shows that \xname{} scales similar to
\msync{} with an increasing number of threads while maintaining a lower overhead
overall. Each thread in this microbenchmark operates on an independent memory
region and mimics a small transactional update by writing to random memory
locations and calling \msync{}. The threads call 500k \msync{}s in total, with
each \msync{} flushing modification from two random writes to their memory
region.}

\ifarxiv
  \subsection{Persistent Memory Applications and Data-Structures}
  We evaluate several applications to show that \xname{} consistently
  outperforms PMDK across various workloads and POSIX \msync on write-heavy
  workloads when running on Intel Optane DC-PMM. Workloads include a linked list
  and a b-tree implementation from Intel's PMDK\ignore{~\cite{pmdk}}, PMDK's
  KV-Store using the YCSB workload\ignore{~\cite{ycsb}}, and Kyoto Cabinet.

\fi

\cutcamera{\boldparagraph{Linked List} \reffig{fig:linkedlist-btree}a shows the performance of a
linked list data structure implemented using PMDK, traditional \msync{} with
4~KiB and 2~MiB page size, \msync{} with data journal, and
\xname{}. \textit{Insert} inserts a new node to the tail of the list,
\textit{Delete} removes a node from the head, and \textit{Traverse} visits every
node and sums up the values. PMDK and \xname{} are crash-consistent, while the
\msync{} implementations are not. Each operation is repeated 1 million times.

Since \xname{} runs the application entirely on DRAM and performs userspace
synchronization with the backing media on a call to \snapshot{}, it
significantly outperforms PMDK on Traverse workload while being competitive in
Insert and Delete. On every call to \msync{}, the traditional filesystem
implementation needs to perform an expensive context switch and TLB
shootdown, slowing it considerably compared to PMDK and \xname{}.\\

\begin{figure}
  \centering
  \begin{subfigure}{\textwidth}
    \includegraphics[width=\linewidth]{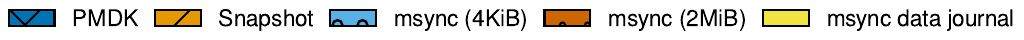}
  \end{subfigure}
  \begin{subfigure}{0.48\textwidth}
    \includegraphics[width=\linewidth]{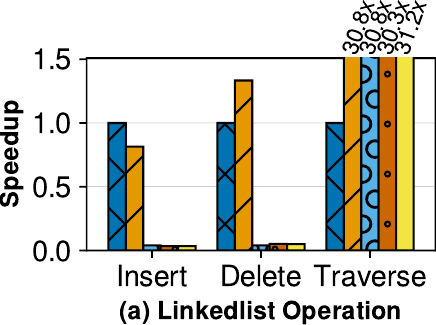}
    \label{fig:linkedlist}
  \end{subfigure}\hfill
  \begin{subfigure}{0.48\textwidth}
    \vspace{3pt}
    \includegraphics[width=\linewidth]{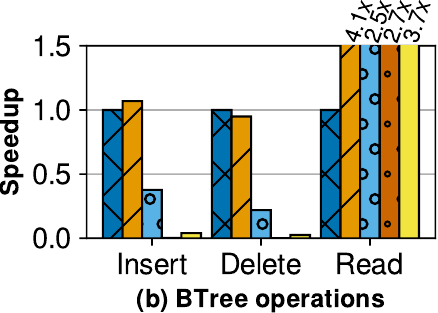}
    \label{fig:map-btree}
  \end{subfigure}
  
  \caption{Performance comparison of PMDK, \xname{}, and \msync{} on Intel
    Optane DC-PMM. Higher is better.}
  \label{fig:linkedlist-btree}
\end{figure}

\boldparagraph{B-Tree} We compare \xname{} against PMDK using a b-tree data
structure of order 8 with 8-byte keys and values on Intel Optane DC-PMM. We use
three workloads: (1) \textit{Insert workload} generates 1 million random 8-byte
keys and values. (2) \textit{Read workload} traverses the tree in depth-first
order. Finally, (3) the \textit{delete workload} deletes all the keys in the
insertion order.

\reffig{fig:linkedlist-btree}b shows that \xname{} performs similarly to PMDK for the
insert and delete workloads while outperforming all \msync{} implementations by
at least \btreeinsdelsnapshotmsyncminspeedup{}. For the read workload, \xname{}
and \msync{} achieve significant speedup (\btreereadpmdkspeeduppmem) over
PMDK. \ignore{However, unlike \msync{},
  \xname{} provides crash-consistency guarantee.}\\
}
\newcommand{\wrkldrot}{\parbox[t]{2mm}{\multirow{7}{*}{\rotatebox[origin=c]{90}{\textbf{Workload}}}}}
\newcommand{\celllinebreak}{\\[-2pt]}

\ifarxiv
\begin{table}
  \caption{{Description of the YCSB workload.}}
  \label{tab:workload-desc}
  \fontsize{7}{10}
  \selectfont
  {}
  \begin{tabular}{|c|c|l|}
      \cline{1-3}
      \wrkldrot{}           & \textbf{A} & Read: 50\%, Update: 50\%          \\\cline{2-3}
                            & \textbf{B} & Read: 95\%, Update: 5\%           \\\cline{2-3}
                            & \textbf{C} & Read: 100\%                       \\\cline{2-3}
                            & \textbf{D} & Insert \& Read latest, delete old \\\cline{2-3}
                            & \textbf{E} & Read-modify-write                 \\\cline{2-3}
                            & \textbf{F} & Short range scans                 \\\cline{2-3}
                            & \textbf{G} & Update: 100\%                     \\\hline

    \end{tabular}
\end{table}
\fi

\cutcamera{\begin{figure}
    \includegraphics[width=\linewidth]{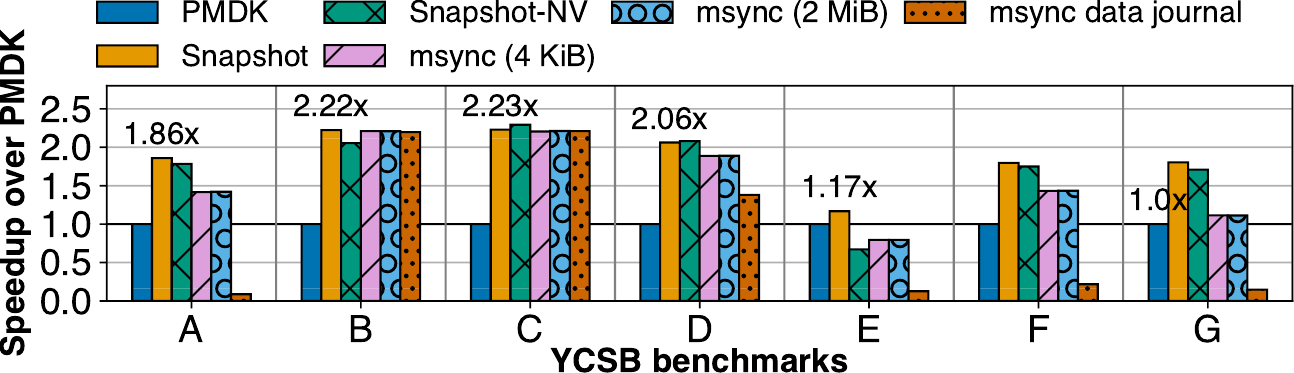}%
    \caption{{KV-store speedup over PMDK with YCSB workload. Higher is
        better. The non-volatile variant uses the undo log to identify
          modified locations. Workload description in
          \reftab{tab:workload-desc}.}}
    \label{fig:simplekv}
  \end{figure}
  }

  \boldparagraph{KV-Store}\label{sec:kv-store-results}
  \ifarxiv{}Next\fi, we compare the performance of \xname{} on
  Optane DC-PMM using a key-value store implemented using a hash table where
  each bucket is a vector. For evaluation, we use the YCSB\ignore{~\cite{ycsb}}
  workloads A-F and an additional write-only workload, G. Each workload performs
  5 million operations on a database with 5 million key-value pairs each.

\ifarxiv\reffig{fig:simplekv}\fi{} shows the
performance of the KV-store against PMDK\cutcamera{\ using different \xname{}
  configurations described in \reftab{tab:configs}}. \cutcamera{For \xname{}, we
  present the results using volatile and non-volatile lists for finding modified
  cachelines on \snapshot{}.} Overall, \xname{} shows between \kvstoreminpmem
and \kvstoremaxpmem performance improvement over PMDK.  \ifarxiv

\else \fi Against \msync{}, \xname{} shows a significant performance
improvement, especially when compared to \msync{} with data journal
enabled. \cutcamera{\xname{} does this while providing an automatic
  crash-consistency guarantee. Moreover, several \xname{} optimizations,
  including using a separate volatile list for tracking dirty data in the memory
  (\refsec{sec:reducing-pm-access}) provide considerable improvement to
  \xname{}'s performance.}

\cutcamera{\begin{figure}
    \includegraphics[width=\linewidth]{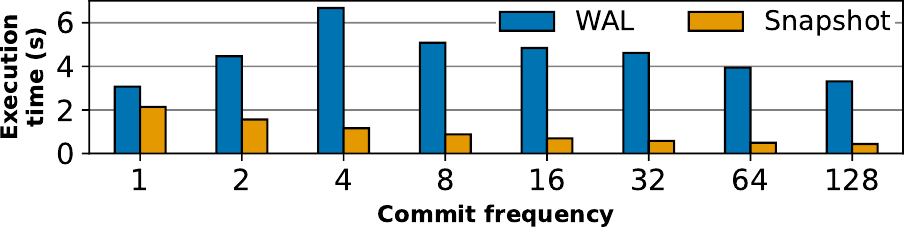}
    {}
  \caption{Performance comparison of commit frequency for writes in Kyoto
    Cabinet on Intel Optane DC-PMM. Average of six runs. Lower is better.}
  \label{fig:kyoto-tx}
\end{figure}}

\boldparagraph{Kyoto Cabinet}
\ifarxiv\reffig{fig:kyoto-tx}\fi{} shows the
\ifarxiv{}performance \fi comparison between Kyoto Cabinet's built-in WAL+\msync{} based
crash-consistency mechanism and \xname{} with a varying number of updates per
transaction. Overall, \xname{} outperforms Kyoto's transaction implementation
by \kyotomin to \kyotomax.

For crash consistency, Kyoto Cabinet combines WAL and \msync{}. For the \xname{}
version, we disable Kyoto Cabinet's WAL implementation. This version, when
compiled with \xname{}'s compiler, is automatically crash-consistent.

\ifarxiv
\begin{figure}
  \begin{minipage}[t]{0.32\linewidth}
    \includegraphics[width=\linewidth]{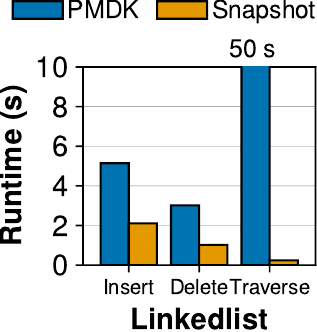}
  \end{minipage}%
  \hfill%
  \begin{minipage}[t]{0.32\linewidth}
    \includegraphics[width=\linewidth]{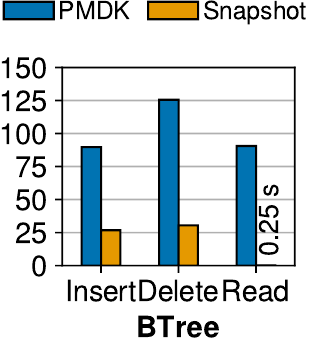}
  \end{minipage}%
  \hfill%
  \begin{minipage}[t]{0.32\linewidth}
    \includegraphics[width=\linewidth]{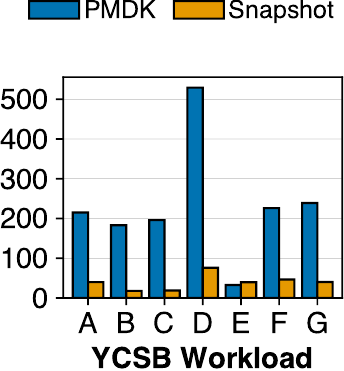} 
  \end{minipage}
  \caption{Linked list, b-tree, and KV-store on Emulated Memory Semantic SSD
    (\refsec{sec:mem-sem-ssd}). Lower is better.}
  \label{fig:mem-sem-results}
  \fixspacing
\end{figure}
\fi

\subsection{CXL-Based Memory Semantic SSDs}
\cutcamera{We evaluate Linked list, B-tree, and KV-Store on our emulated memory semantic
SSD and observe that \xname{} significantly outperforms PMDK for linked list and
b-tree (\reffig{fig:mem-sem-results}).  For linked list,
\xname{} outperforms PMDK by 1.7$\times$, 3.2$\times$, and 171.0$\times$ for
insert, delete, and read, respectively. For b-tree, \xname{} outperforms PMDK by
3.4$\times$, 4.1$\times$, and 364.5$\times$ for insert, delete, and read
workloads, respectively.}

For the KV-Store benchmark\addcamera{\ on our emulated memory semantic SSD},
\xname{} outperforms PMDK \addcamera{(\reffig{fig:cam-ready-results}c)} by up to
10.9$\times$ for all but the `E' workload, where PMDK is 1.23$\times$ faster. As
our emulation is software-based, it does not support the \msync{} system call,
so we did not evaluate Kyoto Cabinet.

\cutcamera{\subsection{Programming effort}
\ifarxiv{}Implementing\fi{} \xname did not require changing
\texttt{boost.interprocess} as any changes to the memory allocator's state are
automatically persisted with the \msync{} call by the workloads. For Kyoto
Cabinet, we changed 11 lines of code, including disabling its \ifarxiv{}built-in\fi{}
crash-consistency mechanism\cutcamera{, demonstrating the utility of \xname{}'s
  simple programming interface for achieving crash consistency}.

\ignore{To understand the programming effort needed to use \xname{} to implement
crash consistency, we looked at the lines of code changed in
\texttt{boost.interprocess}, the shared memory allocator. For
\texttt{boost.interprocess}, we made no changes to its source code to be
crash-consistent. Changes to the memory allocator's state are automatically
persisted with the \msync{} call by the workloads. For Kyoto Cabinet, we changed
11 lines of code, including disabling its built-in crash-consistency
mechanism. Thus, demonstrating the utility of \xname{}'s simple programming
interface for achieving crash consistency.}
}
\cutcamera{
\section{Related Work}
\label{sec:related}

\ignore{\subsection{Persistent memory programming}}

Many prior works have proposed techniques to simplify the persistent memory
programming model. \ignore{\ignore{focused on improving PM programming,
    including finding persistent memory bugs~\cite{liu2020cross,
      neal2020agamotto, liu2021pmfuzz,corundum, Liu:2019:PAF, gorjiara2021jaaru,
      bozdougan2022safepm, gorjiara2022yashme}. Other works have} proposed
  techniques to port existing volatile programs to persistent memory, including
  using deep-learning-assisted transformation~\cite{huang2021ayudante} and
  defining a set of rules for transforming specific classes of programs (for
  example, non-blocking volatile or
  lock-free)~\cite{%
    lee2019recipe, gogte2018sfr}.}
Romulus~\cite{correia2018romulus} uses a twin-copy design, storing both copies
on PM for fast persistence. Romulus uses a redo log to synchronize the active
copy with the backing copy on a transaction commit. Pisces~\cite{gu2019pisces}
is a similar PM transaction implementation that uses dual version concurrency
control (DVCC), where one of the versions is the application data on persistent
memory, and the other is the redo log. \cutcamera{Pisces improves performance by using a
three-stage commit protocol where the stages where the data is durable, visible,
and propagated are decoupled\ignore{, resulting in better scalability and
performance}.} \cutcamera{Pronto~\cite{pronto} introduces the idea of asynchronous
  semantic logging (ASL). Calls to the data structures' public functions in
  Pronto are recorded along with their arguments in a persistent semantic
  log. The semantic log is asynchronously applied to the PM copy using a
  background thread.  Like \xname, in Pronto, the application runs on DRAM and
  gains performance improvement by faster access latency of DRAM.}

\add{Libnvmmio~\cite{libnvmmio} provides an \msync{}-based interface, but it
  does so by intercepting filesystem IO calls, e.g., \texttt{read()} and
  \texttt{write()}. Thus, unlike \xname, Libnvmmio does not support the
  memory-mapped file interface.  Similarly, DudeTM~\cite{dudetm}, while uses
  a memory-mapped interface and stages working copy in the DRAM, requires the
  programmer to use a PMDK-like transactional interface, increasing programming
  effort.}

Automated solutions like a compiler pass to simplify the programming effort
include Atlas~\cite{atlas}, which adds crash-consistency to existing lock-based
programs by using the outermost lock/unlock operations. Synchronization Free
Regions (SFR)~\cite{gogte2018sfr} extend this idea and provide
failure-atomicity between every synchronization primitive and system call.

Similarly, some works use language support to automatically add persistence to
applications written for volatile memory. \ignore{Examples include
  AutoPersist~\cite{autopersist}, go-pmem~\cite{george2020go}, and
  Espresso~\cite{wu2018espresso}. QuickCheck~\cite{shull2019quickcheck}
  speculates whether a Java object will be located on the PM heap and optimizes
  compiler instrumented runtime checks to reduce execution overhead.}
Breeze~\cite{breeze} uses compiler instrumentation for logging updates to PM but
requires the programming to explicitly wrap code regions in transactions and
annotate PM objects. NVTraverse~\cite{friedman200nvtraverse} and
Mirror~\cite{friedman2021mirror} convert lock-free data structures persistent
using programmer annotation and providing special compare and swap operations.

\cutcamera{Memory allocators are important in achieving crash consistency. To
  resume after a crash, persistent memory allocators need to save their metadata
  state along with the allocated data. \ignore{Moreover, persistent memory
    allocators need special care since any bugs or memory leaks in the allocator
    would be permanent.}  Several works in the past have proposed PM
  allocators. Romulus~\cite{correia2018romulus} supports porting any sequential
  memory allocator designed for volatile memory allocation to PM by wrapping all
  the persistent types in a special class that interposes store accesses. This
  is similar to \xname{}'s compiler-based instrumentation, but in contrast to
  Romulus, \xname requires no programmer effort to use a volatile memory
  allocator for \PM.

\ignore{Poseidon~\cite{demeri2020poseidon} provides safety guarantees like
  safeguards against metadata corruption safety and memory leaks.}
LRMalloc~\cite{cai2020understanding} is a PM allocator that persists only
information that is needed to reconstruct the allocator state after a
crash. Metall Allocator~\cite{iwabuchi2019metall} provides a coarse
crash-consistency mechanism by using the underlying DAX filesystem's
copy-on-write mechanism. However, Metall only guarantees persistency when the
Metall allocator's destructor is called, making it impractical for applications
that need higher frequency persistency (e.g., databases).  Kelly et
al.~\cite{pma} present a persistent memory allocator (PMA) to provide a
persistent heap for conventional volatile programs and create a persistent
variant of gawk, \texttt{pm-gawk}~\cite{pm-gawk}.

}

}
\vspace{-2mm}
\section{Conclusion}
\label{sec:conclude}
\vspace{-0.1cm} \xname provides a userspace implementation of failure atomic
\msync{} (\famus{}) that overcomes its performance limitation. \cutcamera{This
  advantage is especially apparent against \msync{} with huge pages enabled.}
\xname's sub-page granularity dirty data tracking based crash consistency
outperforms both per-page tracking of \msync{} and manual annotation-based
transactions of PMDK across several workloads. \cutcamera{Further, \xname{}
  guarantees that the persistent memory state of the application is only updated
  on a call to \msync{}.}

\cutcamera{Further, we study the latency difference between the different ways to write to
persistent memory, NT-Stores vs. cacheline writebacks for uncached data.}

\cutcamera{Finally, }\xname alleviates limitations of \famus{}, enabling applications to
take advantage of faster, byte-addressable storage devices. Moreover, \xname,
unlike \famus{}, completely avoids any system calls for crash consistency or
manual annotation and transactional semantics required by PMDK.

\ifarxiv\else\end{spacing}\fi

 \vspace{-0.15cm}
 \ifarxiv\else\begin{spacing}{0.95}\fi
   {
\small
\bibliography{libpaper/common,libpaper/nvsl,paper}
\bibliographystyle{IEEEtran}
}
   \ifarxiv\else \end{spacing}\fi

\end{document}